\documentclass[preprints,article,accept,moreauthors,pdftex]{Definitions/mdpi}

\pdfoutput=1

\usepackage{subcaption}
\usepackage{amsmath}
\usepackage[toc,page]{appendix}
\usepackage{float}
\usepackage{natbib}
\usepackage{graphicx}
\usepackage{graphicx,subcaption}
\usepackage{ulem}
\usepackage[utf8]{inputenc}
\usepackage{makecell}

\usepackage{array}
\usepackage{wrapfig}
\usepackage{amsmath}
\usepackage{wrapfig}
\usepackage{ulem} 

\firstpage{1} 
\pubvolume{1}
\issuenum{1}
\articlenumber{0}
\pubyear{2022}
\copyrightyear{2022}
\datereceived{} 
\dateaccepted{} 
\datepublished{} 

\hreflink{} 

\usepackage{subcaption}
\usepackage{amsmath}
\usepackage[toc,page]{appendix}
\usepackage{float}
\usepackage{natbib}
\usepackage{graphicx}
\usepackage{graphicx,subcaption}
\usepackage{ulem}
\usepackage[utf8]{inputenc}
\usepackage{makecell}

\usepackage{array}
\usepackage{wrapfig}
\usepackage{amsmath}
\usepackage{wrapfig}
\usepackage{ulem}

\Title{Detection and characterisation of conductive objects using electromagnetic induction and a fluxgate magnetometer}

\TitleCitation{Detection and characterisation of conductive objects using electromagnetic induction and a fluxgate magnetometer}

\Author{Lucy Elson $^{1,\ddagger,}$*\orcidA{}, Adil Meraki $^{1,\ddagger,}$*\orcidD{}, Lucas M. Rushton $^{1}$\orcidE{}, Tadas Pyragius $^{1,2}$\orcidC{} and Kasper Jensen $^{1,}$*\orcidB{}}

\AuthorNames{Lucy Elson, Adil Meraki, Lucas M. Rushton, Tadas Pyragius and Kasper Jensen}

\AuthorCitation{Elson, L.; Meraki, A.; Rushton, L.M.; Pyragius, T.; Jensen, K.}

\address{%
$^{1}$ \quad  School of Physics and Astronomy, University of Nottingham, University Park, Nottingham, NG7 2RD, UK \\
$^{2}$ \quad  Tokamak Energy, 173 Brook Dr, Milton, Abingdon, OX14 4SD, UK}

\corres{Correspondence: lucy.elson@nottingham.ac.uk (LE); merakiadil@gmail.com (AM); kasper.jensen@nottingham.ac.uk (KJ)}

\secondnote{These authors contributed equally to this work.}

\abstract{Eddy currents induced in electrically conductive objects can be used to locate metallic objects as well as to assess the properties of materials non-destructively without physical contact. This technique is useful for material identification, such as measuring conductivity and for discriminating whether a sample is magnetic or non-magnetic. In this study, we carried out experiments and numerical simulations for the evaluation of conductive objects. We investigated the frequency dependence of the secondary magnetic field generated by induced eddy currents when a conductive object is placed in a primary oscillating magnetic field.   According to the electromagnetic theory, conductive objects have different responses at different frequencies. 
Using a table-top setup consisting of a fluxgate magnetometer and a primary coil generating a magnetic field with frequency up to 1~kHz, we are able to detect aluminium and steel cylinders using the principle of electromagnetic induction. The experimental results are compared with numerical simulations and we find overall a good agreement. This technique enables identification and characterisation of objects using their electrical conductivity and magnetic permeability.}

\keyword{Electromagnetic induction; conductivity; eddy current; magnetic field; non-destructive testing; numerical simulation; magnetic permeability; fluxgate magnetometer} 

\begin{document}

\section{Introduction}
Electromagnetic induction is routinely used in eddy current testing as a non-destructive technique for flaw detection and material characterisation \cite{Martin2011,s20092608, abdalla2019challenges}. This technique offers the advantage of non-contact scanning without causing damage to the sample under test. Such measurements have various applications - for example in the detection, localisation and characterisation of metallic objects in the defence, aerospace and quality control industries \cite{4526904,92841,10.1117/12.163849, SHAKOOR20111736,ALMEIDA2013359}. The method is based on detecting and characterising electrically conductive objects using an active excitation, where an oscillating primary magnetic field $\textbf{B}_{1}(t)$ created by a coil induces eddy currents in the object. The eddy currents then create a secondary magnetic field $\textbf{B}_{\text{ec}}(t)$ which can be measured by a sensitive magnetometer, such as a fluxgate magnetometer. This technique can be used to detect a wide range of objects, as it is sensitive to both the electrical conductivity $\sigma$ and the magnetic permeability $\mu= \mu_0\mu_r$ of the object, where $\mu_0$ is the vacuum permeability and $\mu_r$ the relative permeability. It has also been shown that measuring both the amplitude and phase of the magnetic field can be used to reconstruct the eddy currents. This principle finds applications in various areas, such as in the monitoring of fuel cells \cite{rasson2006geomagnetics,app11073069}.

A main challenge when detecting a metallic object is discriminating the object, such as an unexploded ordnance UXO, from the noisy environment it is in \cite{billings2004discrimination}. It takes time and resources to identify the object, especially due to false signals from other metal objects and cultural features such as metal buildings, pipelines and oil well casings. By measuring the secondary magnetic field of an electrically conductive object which is placed in a low frequency primary magnetic field, distinct spectral characteristics such as electrical conductivity, magnetic permeability, object geometry and size can be obtained \cite{1192099,917876,s21041092}.   

In this work, we have built a table-top setup with coils and a commercial fluxgate magnetometer. With this setup, we carried out a systematic study in which a number of metallic objects are detected at different positions and with different excitation frequencies. From the frequency dependence of the measured induced magnetic field, we extracted values for the electrical conductivities and magnetic permeabilities of the objects by fitting experimental data to analytical formulae. In order to validate our experimental results we have built a range of different COMSOL models and have made comparisons between the experimental results and numerical simulations. The results are in good agreement with the numerical simulations performed in COMSOL.

This work is organised as follows: first, we describe the experimental setup and methods with metallic objects placed in different configurations. This includes on-axis eddy current measurements for (a) varying the frequency of the primary oscillating magnetic field and (b) varying the position of the object along the $z$-axis relative to the excitation coil and magnetometer. We then present off-axis measurements where the primary coil and magnetometer are fixed in position, but the object is moved off-axis (i.e. along the $y$-axis at a fixed $z$-position). We present results for solid and hollow cylinders made of aluminium and steel.

\section{Setup and Methods} \label{method}

\begin{figure}[t]
\centering
\includegraphics[width=0.8\linewidth]{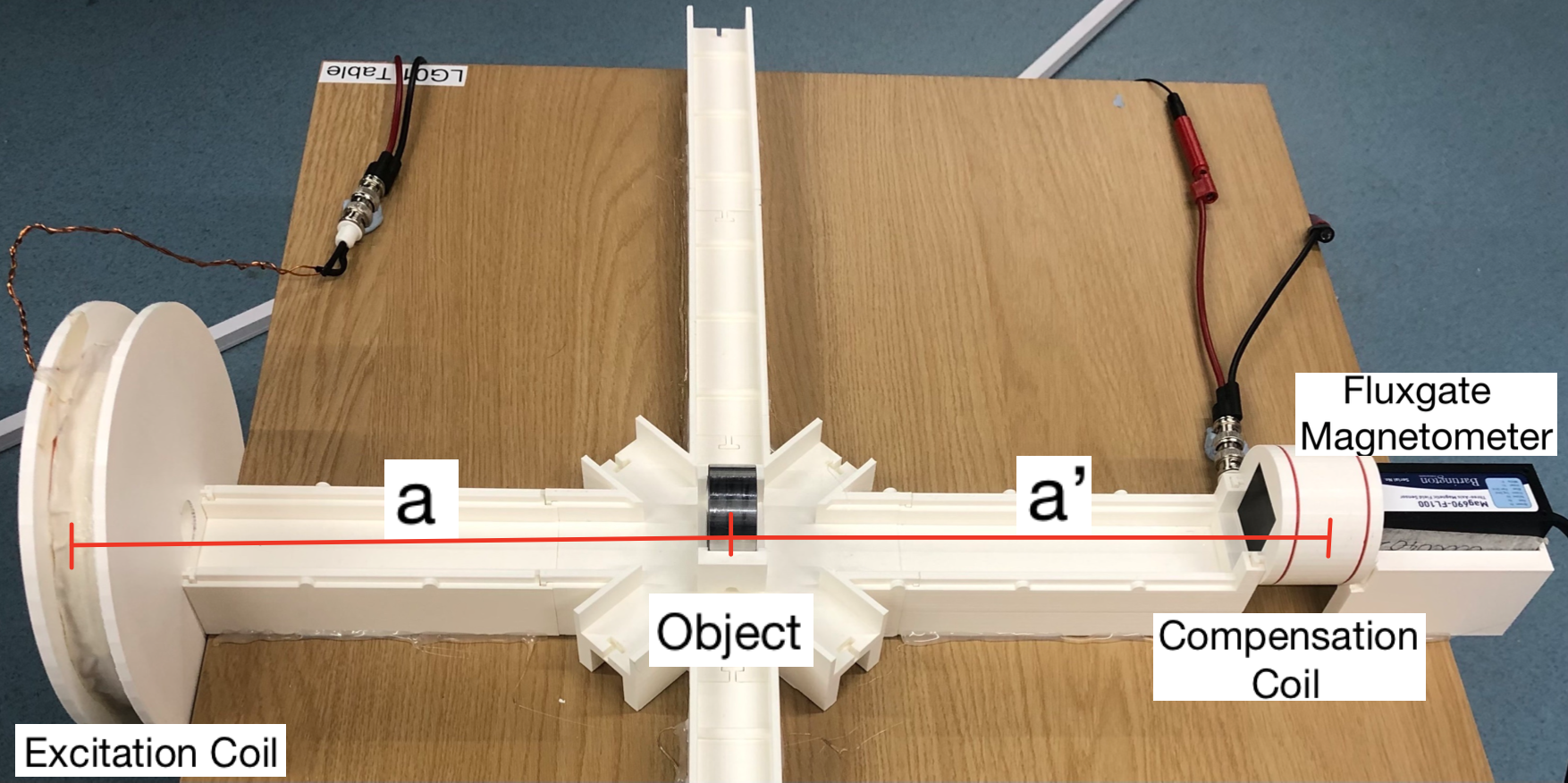}
\caption{Table-top active detection system consisting of an excitation coil, a fluxgate magnetometer, a compensation coil, and an object which can be placed either on-axis or off-axis.}
\label{fig:Lucy_on_set_up}
\end{figure}

Our table-top setup for detecting and characterising metallic samples is shown in Fig.~\ref{fig:Lucy_on_set_up}. The setup is 3D-printed which allows for components to be placed with high precision.
The experiments are controlled with a sbRIO-9627 field-programmable gate array (FPGA) which is programmed in LabVIEW. The FPGA can output sinusoidal signals, record data, perform lock-in amplification, apply real-time feedback, and analyse data. Magnetic fields are detected with a Bartington MAG690 fluxgate magnetometer, which has a scale factor of 100~mV/$\mu$T  and a bandwidth of $\approx 1$~kHz. The fluxgate magnetometer measures all three components of the magnetic field, but for simplicity only the $z$-component is recorded.
In the experiments the data is taken with the  magnetic field oscillating at a particular frequency $\nu$. The resulting oscillating signal $S(t) =  R \cos \left(2\pi \nu t + \phi \right) 
= I \cos \left(2\pi \nu t \right) - Q \sin \left(2\pi \nu t \right)$ can be decomposed into the in-phase $I$ and out-of-phase $Q$ components. These components are detected with a lock-in amplifier implemented on the FPGA. 

Two coils, an excitation coil and a compensation coil, are used for generating magnetic fields. 
The excitation coil produces a primary field $\mathbf{B}_1(t)$ oscillating at a particular frequency ranging between 10 - 1000~Hz. 
The reference phase of the lock-in amplifier was adjusted such that the primary field was detected in the in-phase component $I$ only. A compensation coil was used - it has a one-turn Helmholtz configuration with a radius of 3~cm placed around the magnetometer's detection point. This creates an additional magnetic field $\mathbf{B}_2(t)$, the `compensation field', that is at at the same frequency as the primary field. It cancels the primary magnetic field at the position of the magnetometer, such that the total magnetic field $\mathbf{B}_1(t) + \mathbf{B}_2(t) \approx 0$ in the absence of an object.

The excitation coil has an 8~cm radius, 60~windings and is positioned such that the centre of the coil is 48.4~cm away from the detection point of the magnetometer. In order to produce a magnetic field, a sinusoidal voltage of $\approx$~7.2~V is sent to the excitation coil and the phase is adjusted such that the signal is in the in-phase component of the lock-in output. The applied voltage generates a current of $I_{e} = 0.53$~A in the excitation coil corresponding to a magnetic dipole moment of $\mu_{e} = 0.64$~Am$^{2}$ (pointing in the $z$-direction). The magnetic field produced is $B_1 = 1.09~\mu$T at the position of the magnetometer. To compensate for this field a $0.24$~V oscillating voltage is applied to the compensation coil. This generates a current of $I_{c} = 0.036$~A in the compensation coil corresponding to a magnetic dipole moment of $\mu_{c} = 1.02 \times 10^{-4}$~Am$^2$. 

When a conductive object is placed between the compensation coil and the excitation coil, eddy currents are induced in the object, producing a secondary magnetic field (or an 'induced field') $\textbf{B}_{\text{ec}}(t)$ oscillating at the same frequency as the primary field. Note that the eddy currents are mainly generated by the primary field as the compensation field will be small (compared to the primary field) at the position of the object. The amplitude of the secondary field is therefore proportional to the amplitude of the primary field $|B_{\mathrm{ec}}|\propto|B_1|$. 
Due to the applied compensation field, the magnetometer directly measures the secondary field as the total oscillating field $\textbf{B}_{\text{tot}}(t)=\mathbf{B}_1(t) + \mathbf{B}_2(t)+\textbf{B}_{\text{ec}}(t) \approx\textbf{B}_{\text{ec}}(t)$ at the magnetometer position. 
Applying a compensation field in order to measure the secondary field directly can be convenient and if using an optically pumped magnetometer for detecting the magnetic field, the signal-to-noise ratio of the measurement can improve by several orders of magnitude \cite{Jensen2019,Deans2020apl,rushton2022unshielded}. However, we note that the stability and noise in our measurements with the fluxgate magnetometer was independent of whether the compensation field was applied or not (see Appendix~\ref{appendix:Allan}).

\begin{figure}[H]
\centering
\includegraphics[width=0.45\linewidth]{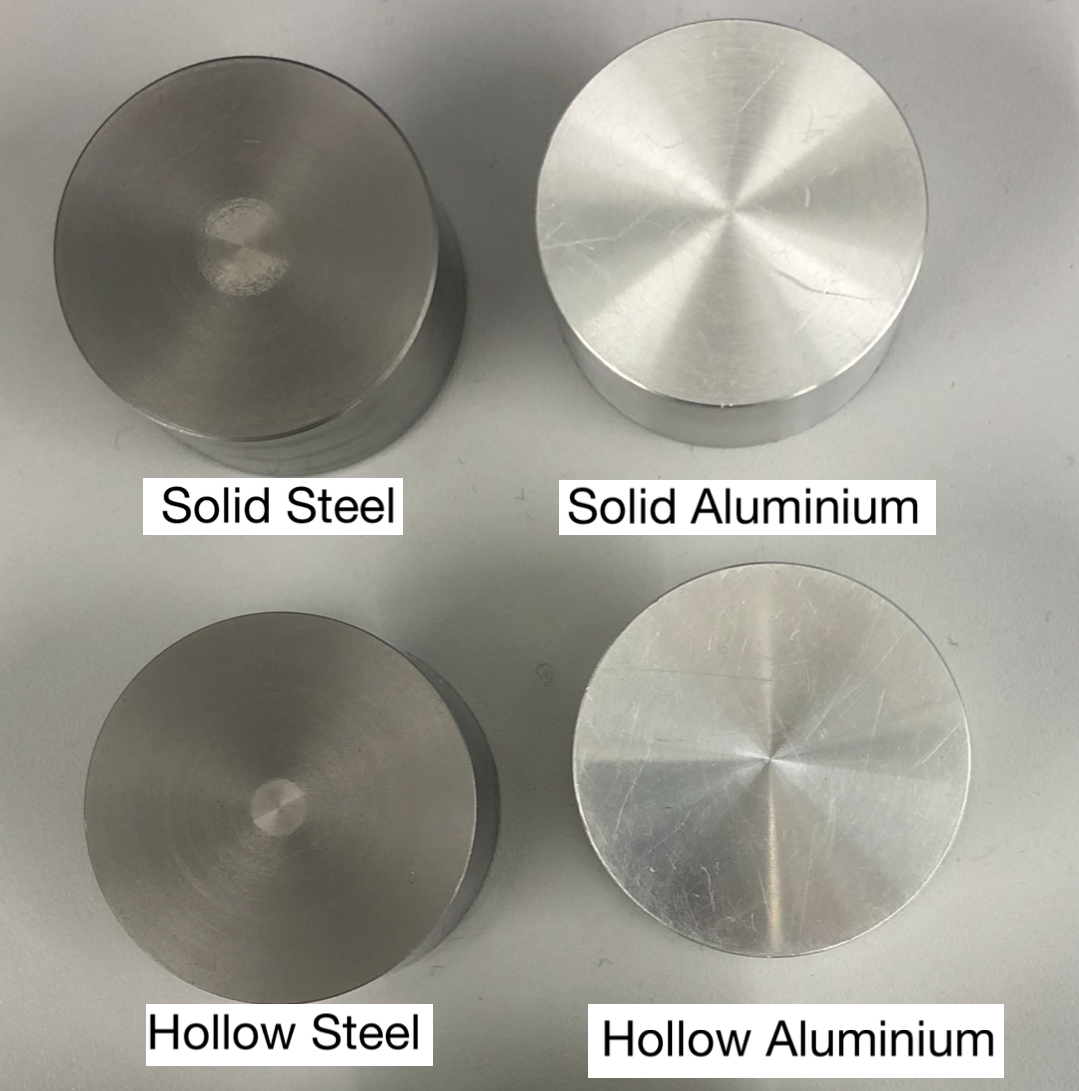}
\caption{6061 T6 aluminium and 440c steel samples.}
\label{fig:Lucy_samples}
\end{figure}

We detect and characterise four different samples (see Fig.~\ref{fig:Lucy_samples}).
The samples used were solid/hollow cylinders with radii of 2~cm and a width of 2~cm. The hollow cylinders have a thickness of 4~mm. The cylinders are made of either 6061 T6 aluminium or 440c steel. 6061 T6 aluminium has an electrical conductivity of $\sigma = 24.6$~MS/m, is non-magnetic and has a relative magnetic permeability  $\mu_{r}=1$ \cite{davis1993aluminum}. 440c steel has an unknown relative permeability, according to its data sheet \cite{davis1994stainless}. The electrical conductivity is unknown but can be determined experimentally \cite{honke2018metallic}. 

Using our table-top setup, we measure how the secondary field  depends on the frequency of the primary field and on the distance from the excitation coil to the sample (and hence from the sample to the fluxgate). For on-axis measurements the object is placed directly between the excitation coil and the magnetometer. In order to study how varying the frequency affects the induced eddy currents, the sample is placed 22.4~cm away from the front of the excitation coil and the frequency is varied between 10~Hz and 1~kHz. When varying the distance of the object a constant frequency of 500~Hz is used. The conductive objects are placed in approximately 5~cm intervals, beginning at 5~cm from the front of the excitation coil to 39.5~cm away. The off-axis measurements are done with the samples being approximately half-way between the two coils, 22.4~cm away from the front of the excitation coil. The conductive objects are placed from 0~cm to 34~cm off-axis and the induced magnetic field is measured.


\section{Numerical Simulations}

\begin{figure}[H]
\centering
\begin{subfigure}[b]{0.49\linewidth}
\centering 
\includegraphics[width=\linewidth]{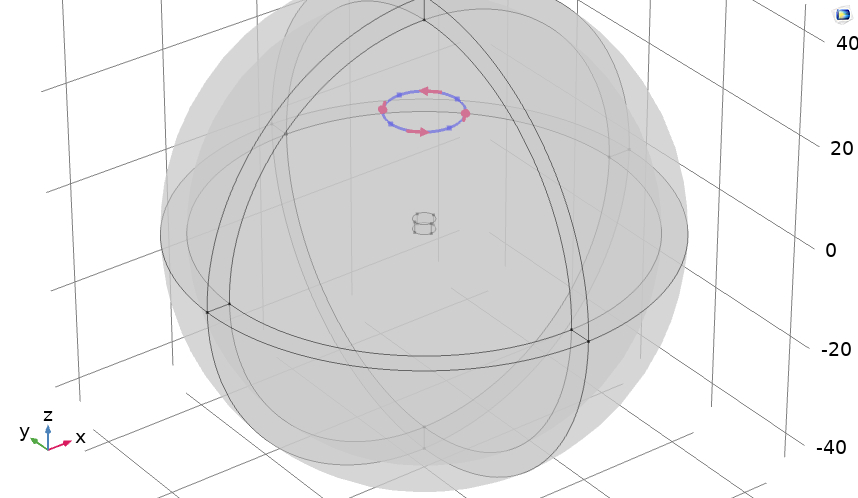}
\caption{}
\label{fig:Model1}
\end{subfigure}
\centering
\begin{subfigure}[b]{0.49\linewidth}
\centering
\includegraphics[width=\linewidth]{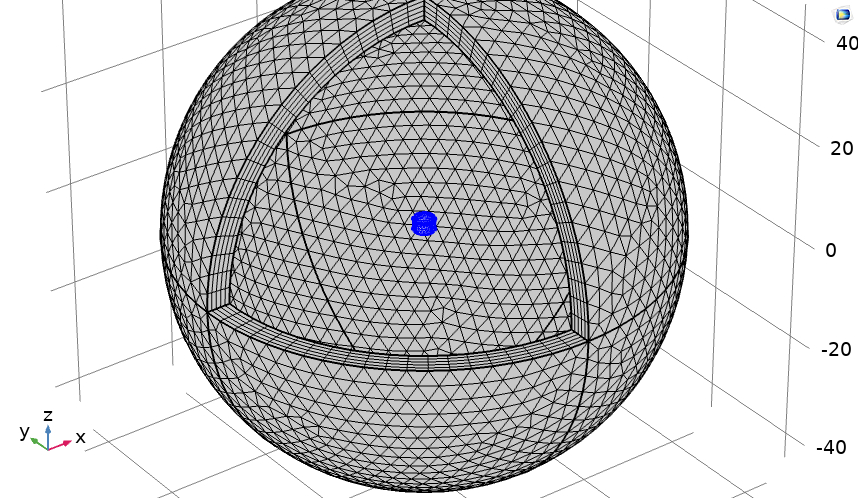}
\caption{}
\label{fig:Model_Mesh}
\end{subfigure}
\caption{Diagrams of the 3D finite element model (COMSOL) showing (a) an object at the origin, the primary coil and the boundary layer and (b) the free tetrahedral elements for the object and the finite domain that was employed in this study.} 
\label{fig:Model}
\end{figure}

In this study eddy current simulations are performed in COMSOL Multiphysics 5.6 using the AC/DC module. The experimental setup is built as a 3D model (Fig.~\ref{fig:Model}). The model consists of a circular coil placed above a metallic object. To reduce complexity, an imaginary single-turn coil is chosen for the primary magnetic field. The coil and the object are placed in the finite sphere air domain whose size is 10 times bigger than the size of the object. As seen in Fig.~\ref{fig:Model} the model also includes the infinite element domain, which is one-tenth of the overall dimension of the model. The functionality of the infinite domain means that the governing equations behave similarly to nature and achieve a non-reflecting boundary condition. The finite element mesh is used to subdivide the CAD model into smaller domains, where a set of equations are solved. As these elements are made as small as possible (the mesh is refined), the solution will approach the true solution. Figure~\ref{fig:Model_Mesh} shows that the finite element mesh consists of three-dimensional tetrahedral solid elements, and 5 layers of infinite element meshes which have been added to the spherical domain. All of the simulations were performed on a workstation using a 3.60 GHz Intel(R) Xeon processor with a 128 GB RAM.

Each simulation was run twice, first with the object present (matching the properties of those used experimentally) and then secondly without the object present. Instead of removing the object from the simulation, its properties (most notably its electrical conductivity and magnetic permeability) were changed to match that of the host medium (air). By using this technique the mesh is preserved in both cases and hence the influence of the mesh on the results is eliminated. The difference between these two simulation outputs is the magnetic field induced in the object. 

\begin{figure}[H]
\centering
\begin{subfigure}[b]{0.49\linewidth}
\centering
\includegraphics[width=\linewidth]{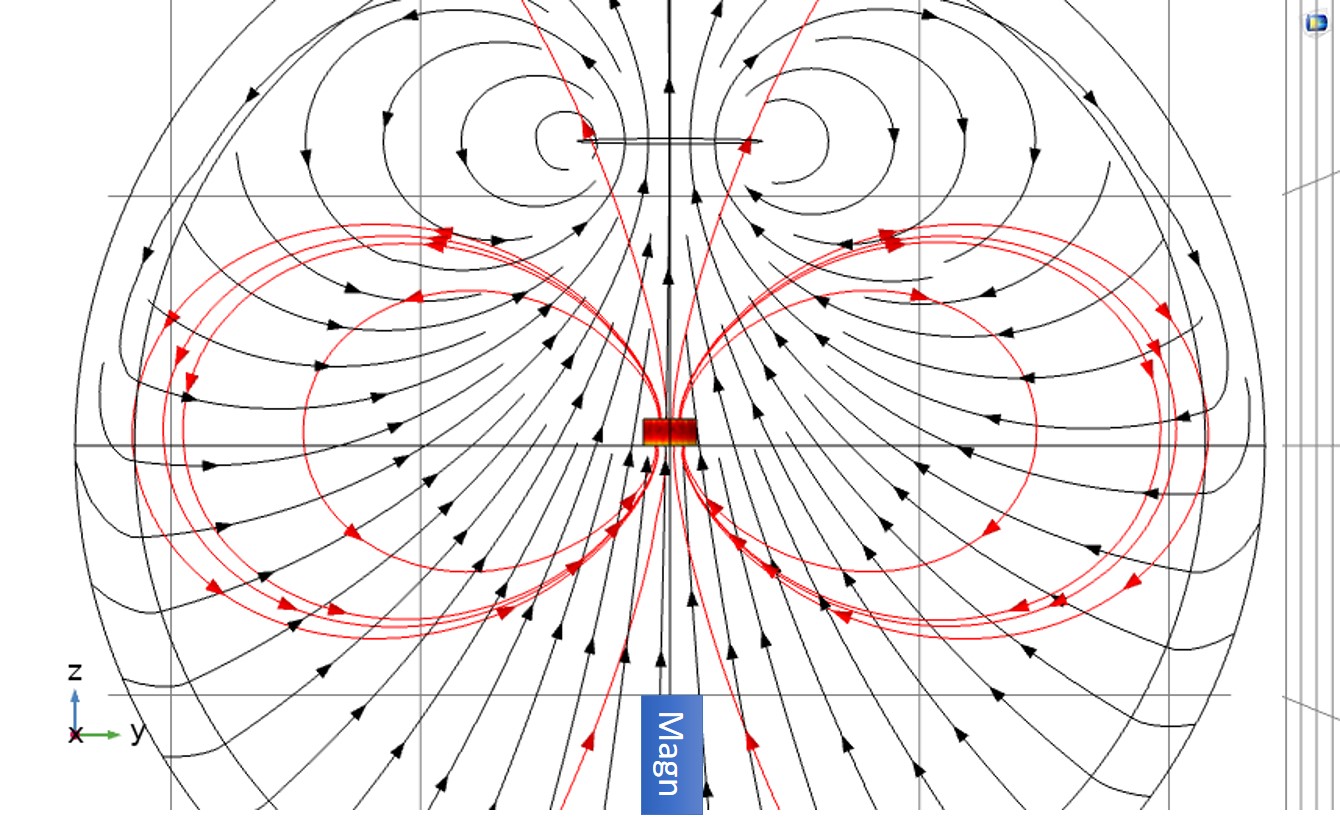}
\caption{}
\label{fig:eddy_current_1}
\end{subfigure}
\centering
\begin{subfigure}[b]{0.49\linewidth}
\centering
\includegraphics[width=\linewidth]{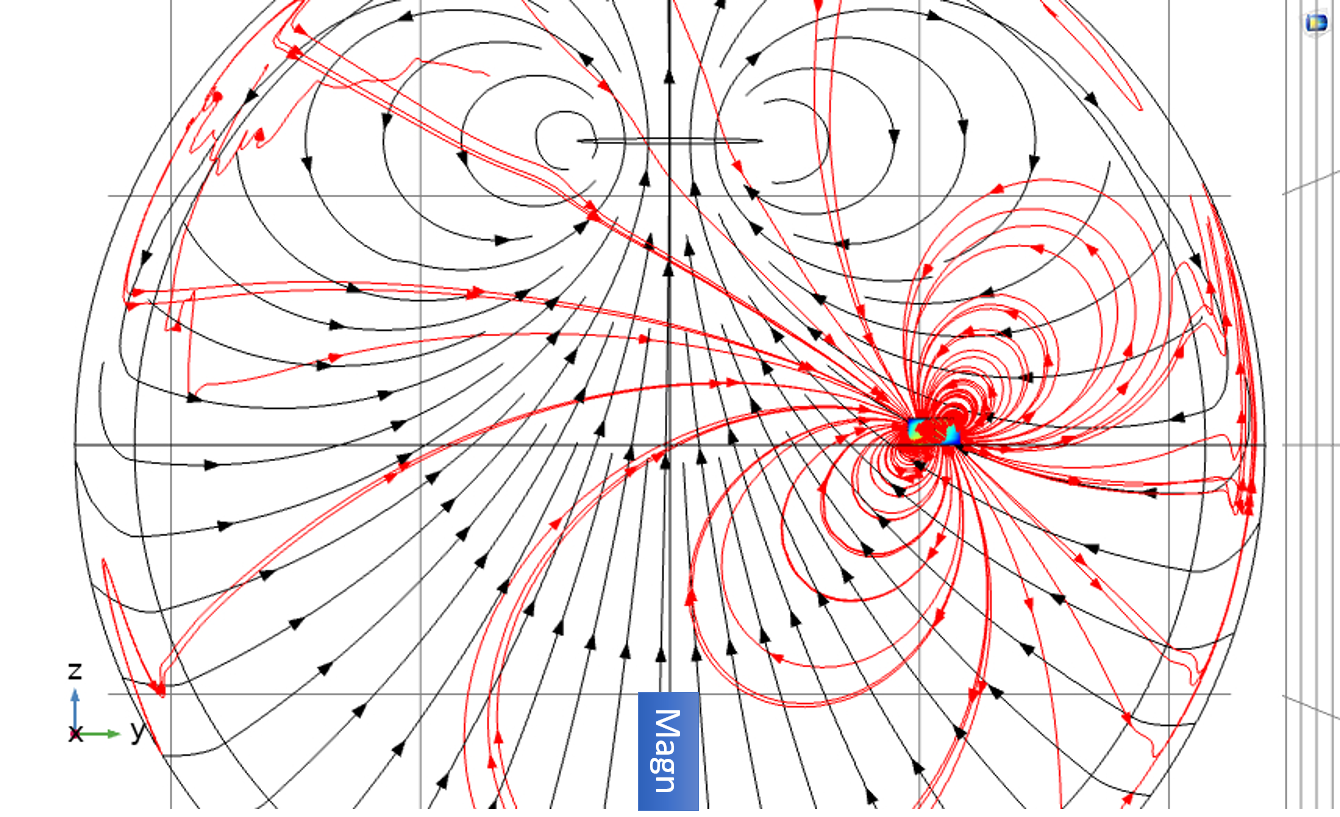}
\caption{}
\label{fig:eddy_current_2}
\end{subfigure}
\caption{Simulation model results when a solid metallic cylinder is placed (a) on-axis and (b) off-axis. The magnetic moment of the primary coil (black line) is aligned along the $z$-axis and stream plots of the magnetic filed lines (black lines with arrows) produced by the primary coil are shown. The induced secondary magnetic field $\textbf{B}_{\text{ec}}(t)$ generated by eddy currents in the object are depicted with the red lines and arrows.}
\label{fig:eddy_current}
\end{figure}

Figure~\ref{fig:eddy_current} shows the directions of the primary and secondary magnetic fields when an object is placed on-axis and off-axis, respectively. When the object is placed on-axis, the primary and secondary magnetic fields only have a $z$-component at the magnetometer position. When the object is placed off-axis in the $y$-$z$-plane, the secondary field will in general have both $y$- and $z$-components. In the following section we will present experimentally measured values for the $z$-component of the secondary field and compare those to values found from numerical simulations.

\section{Results and Discussion}

\begin{figure}[H]
\centering
\begin{subfigure}[b]{0.49\linewidth}
\centering
\includegraphics[width=\linewidth]{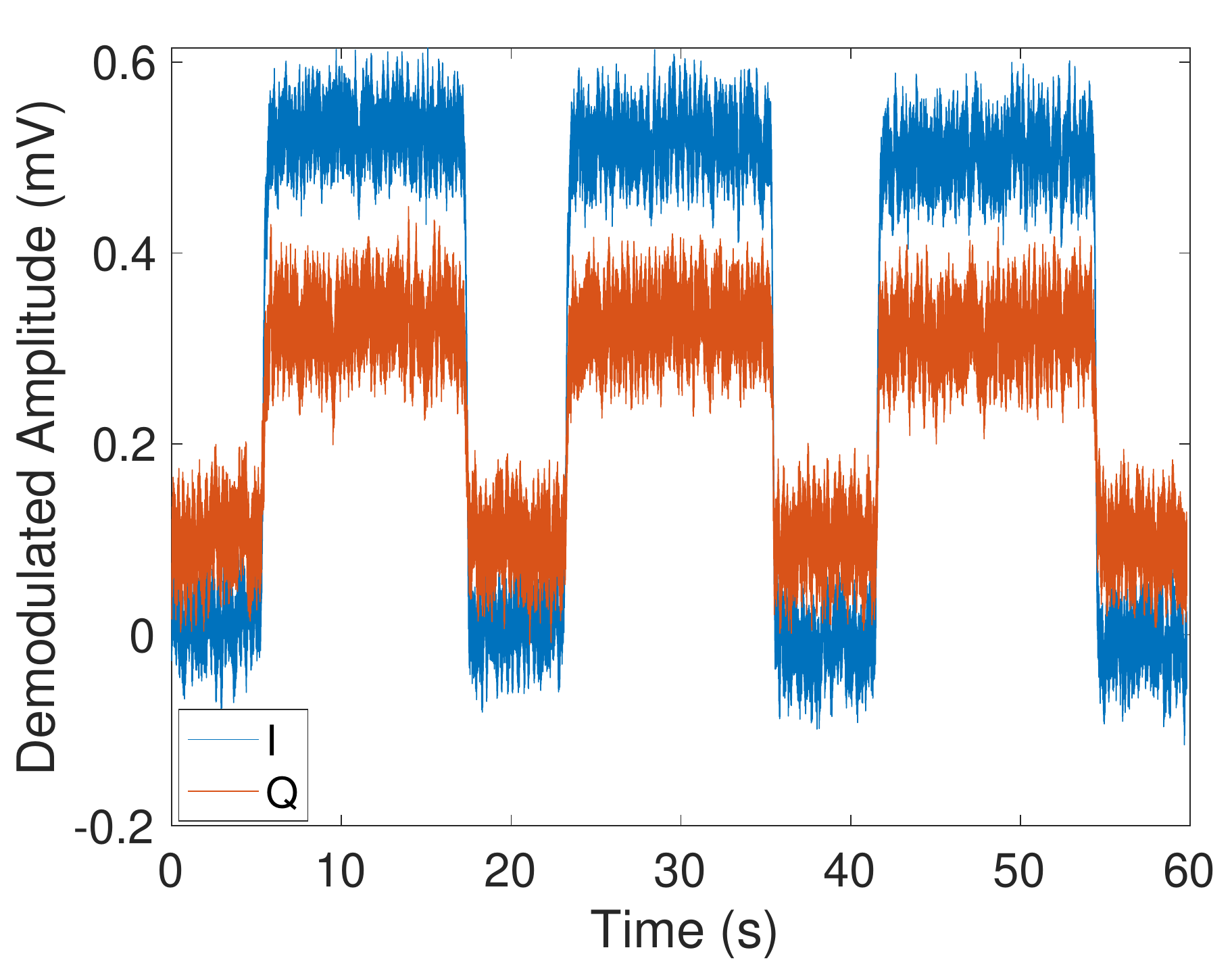}
\caption{}
\label{fig:Lucy_Solid_Alu_time_trace}
\end{subfigure}
\centering
\begin{subfigure}[b]{0.49\linewidth}
\centering
\includegraphics[width=\linewidth]{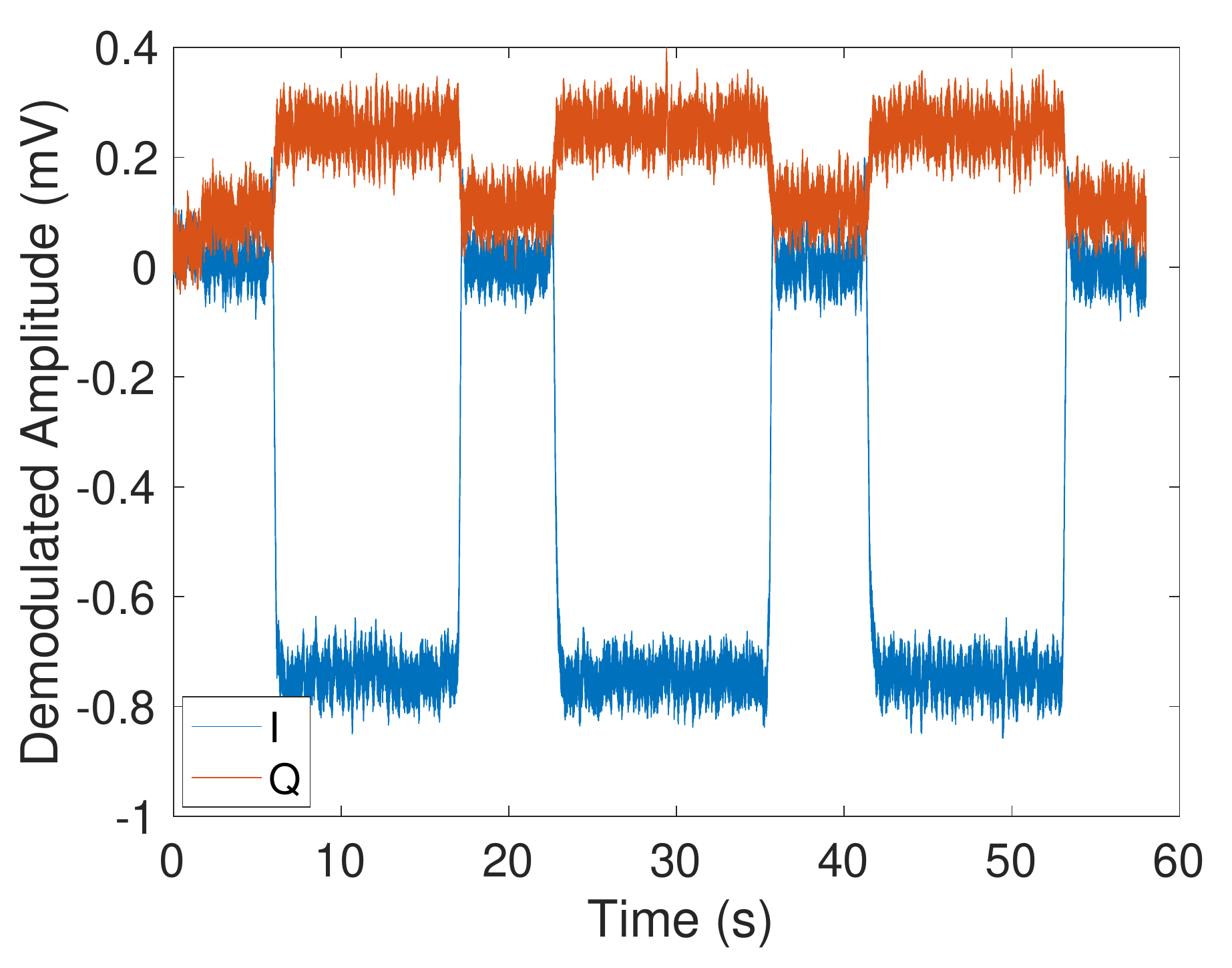}
\caption{}
\label{fig:Lucy_Solid_Steel_time_trace}
\end{subfigure}
\caption{Time traces of the lock-in outputs when detecting conductive objects with the fluxgate magnetometer and the active detection setup for (a) the solid aluminium cylinder and (b) the solid steel cylinder. The conductive object is placed 22.4 cm away from the excitation coil. For these measurements the primary magnetic field is $1.09~\mu$T at the magnetometer position and oscillating at a frequency of 500 Hz.} 
\label{fig:_Lucy_time_trace}
\end{figure}

Figure~\ref{fig:_Lucy_time_trace} shows examples of experimentally recorded time traces when conductive objects were placed into the setup for $\sim 10$~seconds and then taken back out for $\sim 5$~seconds. Three repeats of this measurement were taken. The time traces show the demodulated fluxgate magnetometer signals $I$ and $Q$ for an excitation frequency of $500$~Hz when aluminium and steel cylinders were placed into the setup, respectively. From such time traces the in-phase $\Delta I$ and out-of-phase $\Delta Q$ components of the secondary magnetic field can be found by subtracting the signals with and without the object. This is then used to calculate the magnitude of the secondary field relative to the primary magnetic field $|B_{\text{ec}}|/|B_{1}|$ measured at the magnetometer position, and the phase $\phi$ of the secondary magnetic field with respect to the primary magnetic field. To take the bandwidth of the magnetometer into account, the primary field was recorded at every frequency. 
Time traces were taken for a range of frequencies between $10$~Hz and $1$~kHz and the results are shown in Figure~\ref{fig:Lucy_Freq}. 

\subsection{Varying Frequency}\label{freq}

\begin{figure}[H]
\centering
\begin{subfigure}[b]{0.49\linewidth}
\centering
\includegraphics[width=\linewidth]{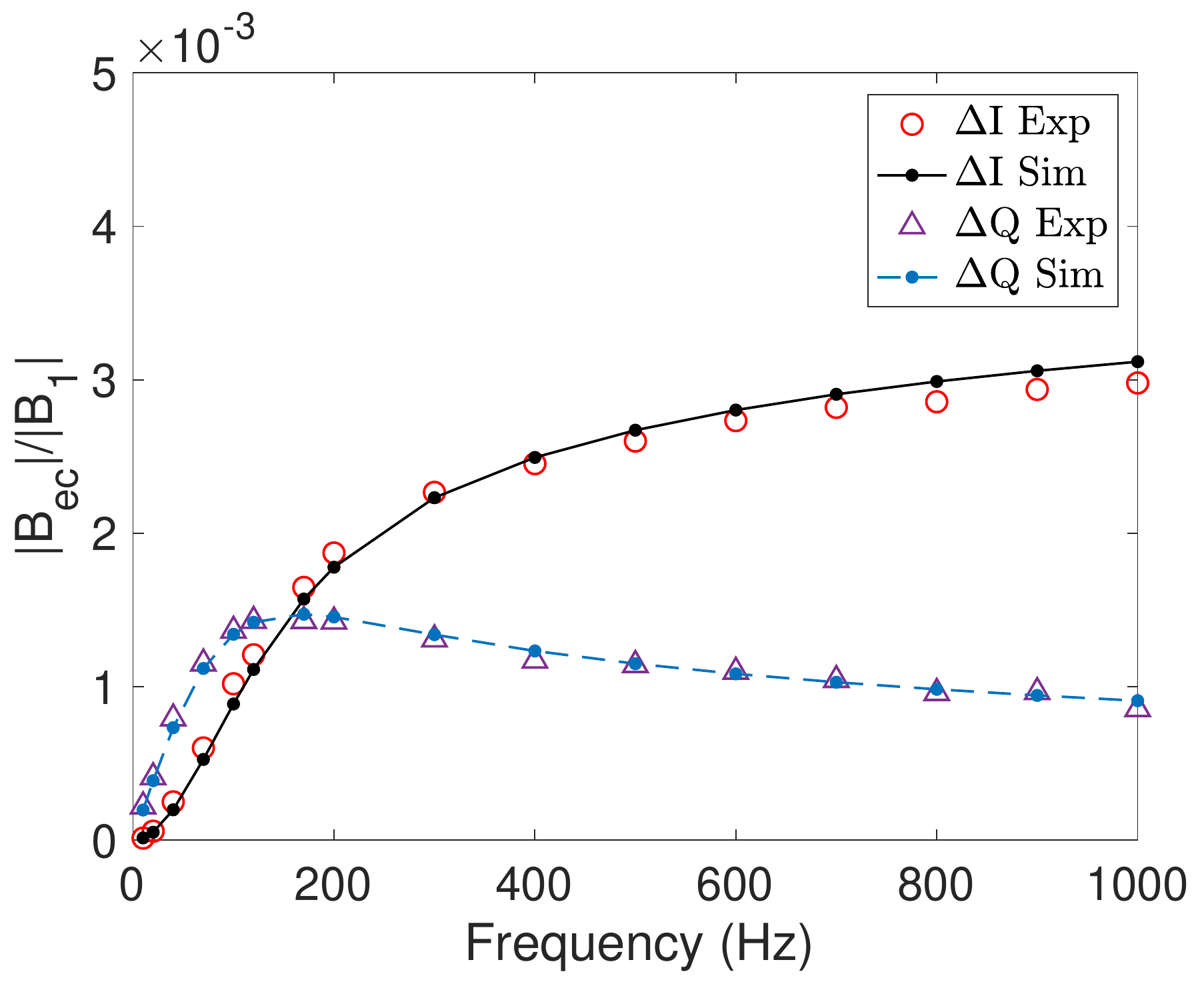}
\caption{}
\label{fig:Lucy_Freq_Solid_Alu_I/Q}
\end{subfigure}
\centering
\begin{subfigure}[b]{0.49\linewidth}
\centering
\includegraphics[width=\linewidth]{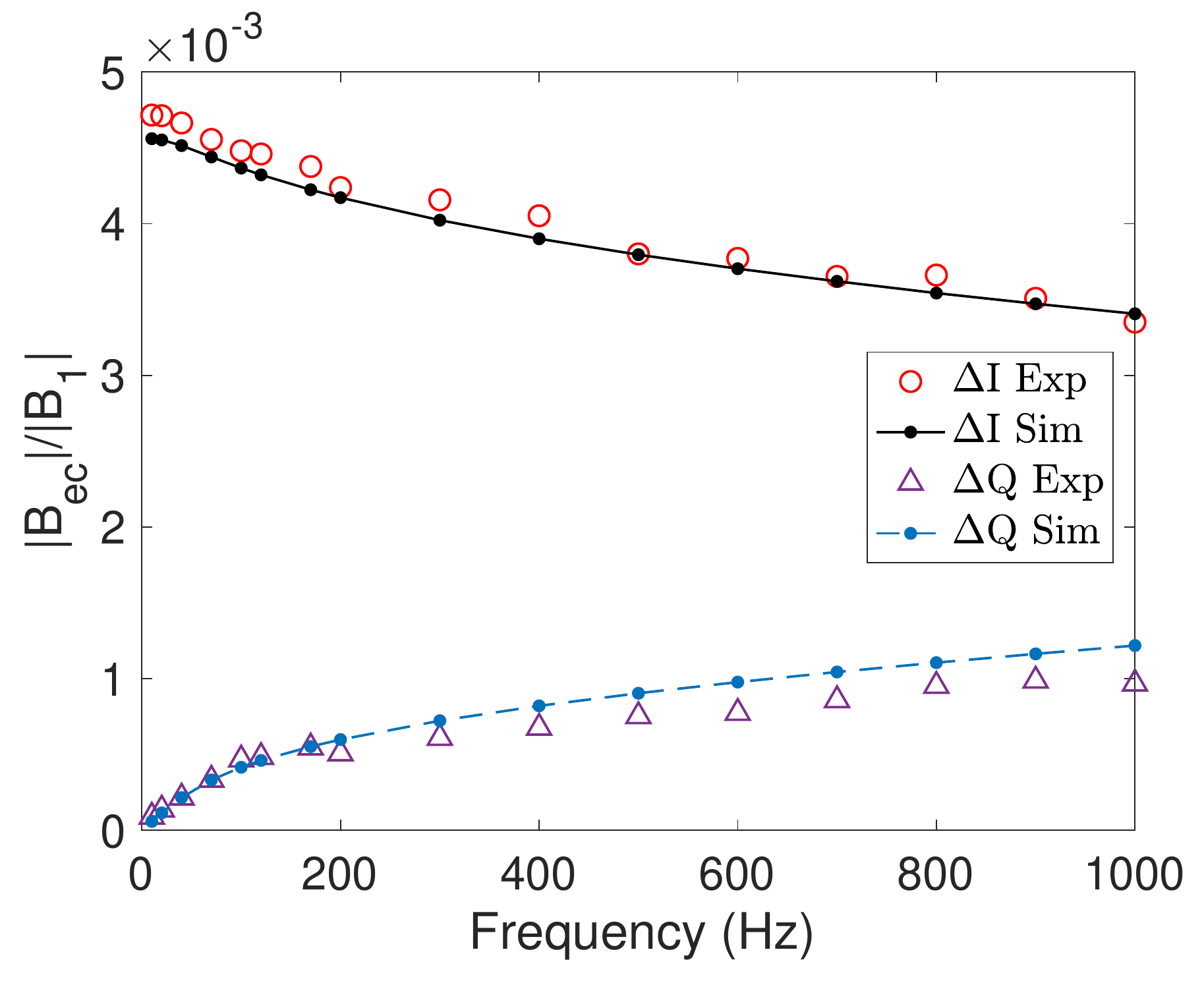}
\caption{}
\label{fig:Lucy_Freq_Solid_Steel_I/Q}
\end{subfigure}
\centering
\begin{subfigure}[b]{0.49\linewidth}
\vspace{0.2cm}
\centering
\includegraphics[width=\linewidth]{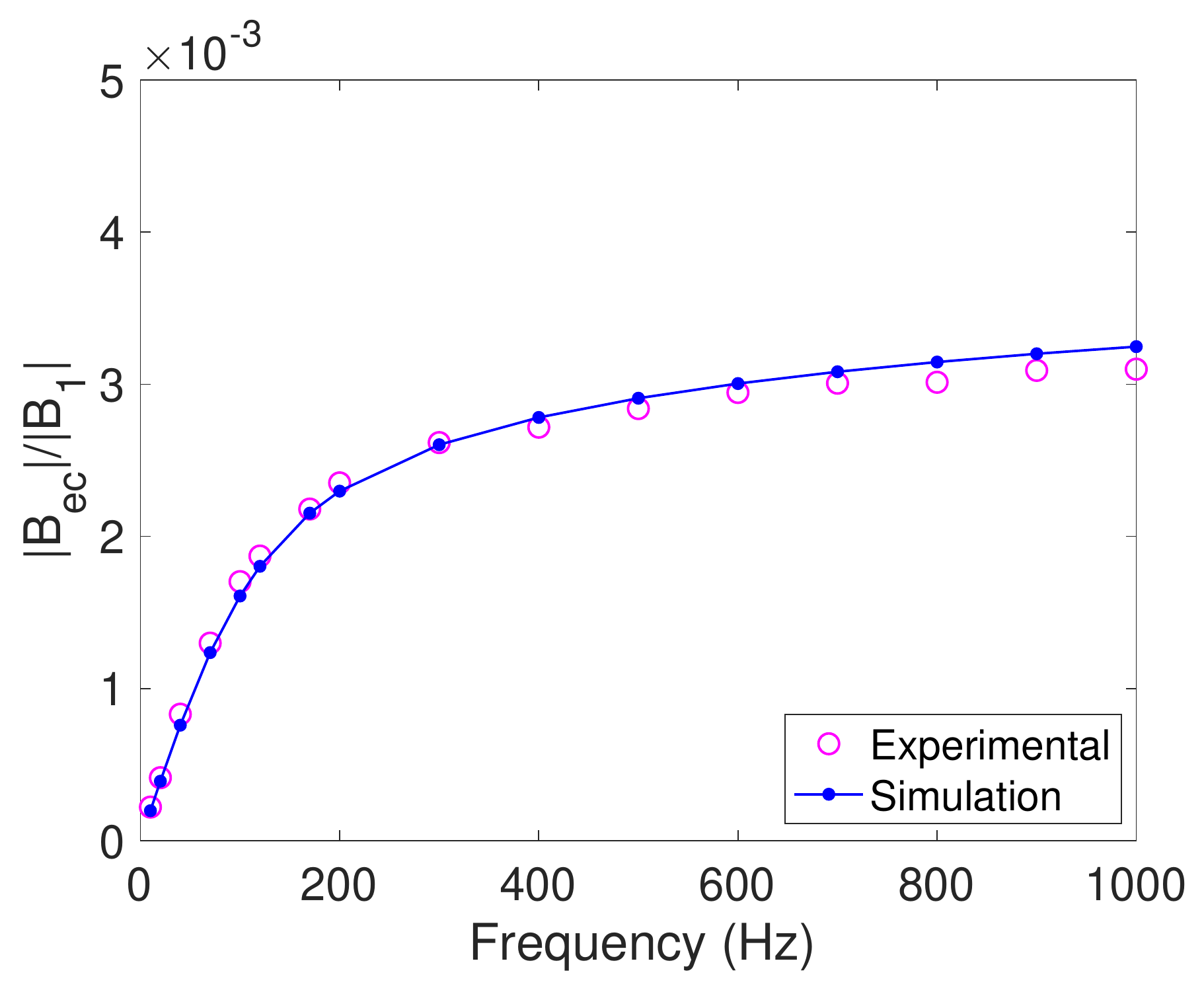}
\caption{}
\label{fig:Lucy_Freq_Solid_Alu_B_ratio}
\end{subfigure}
\centering
\begin{subfigure}[b]{0.49\linewidth}
\centering
\includegraphics[width=\linewidth]{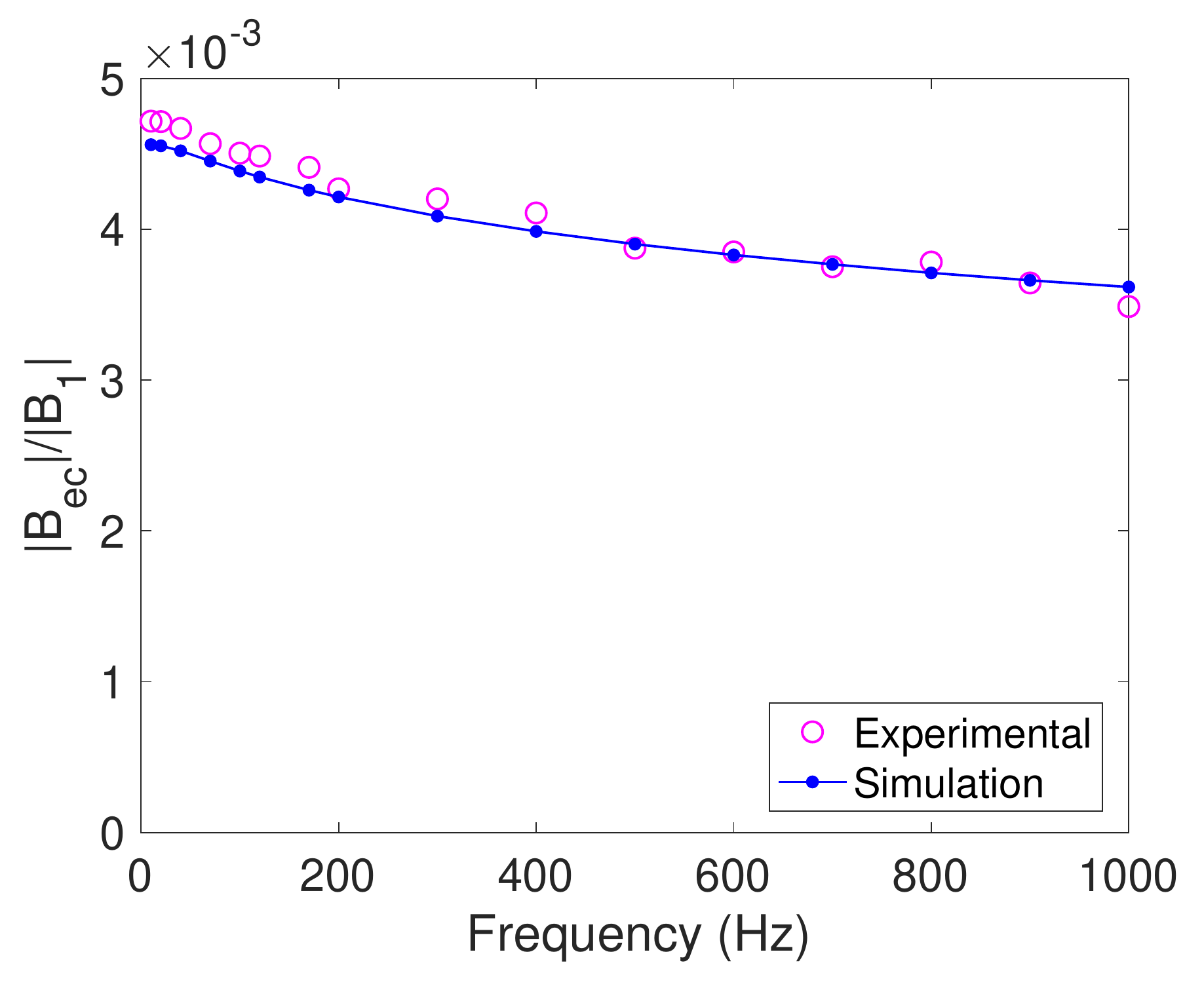}
\caption{}
\label{fig:Lucy_Freq_Solid_Steel_B_ratio}
\end{subfigure}
\centering
\begin{subfigure}[b]{0.49\linewidth}
\vspace{0.2cm}
\centering
\includegraphics[width=\linewidth]{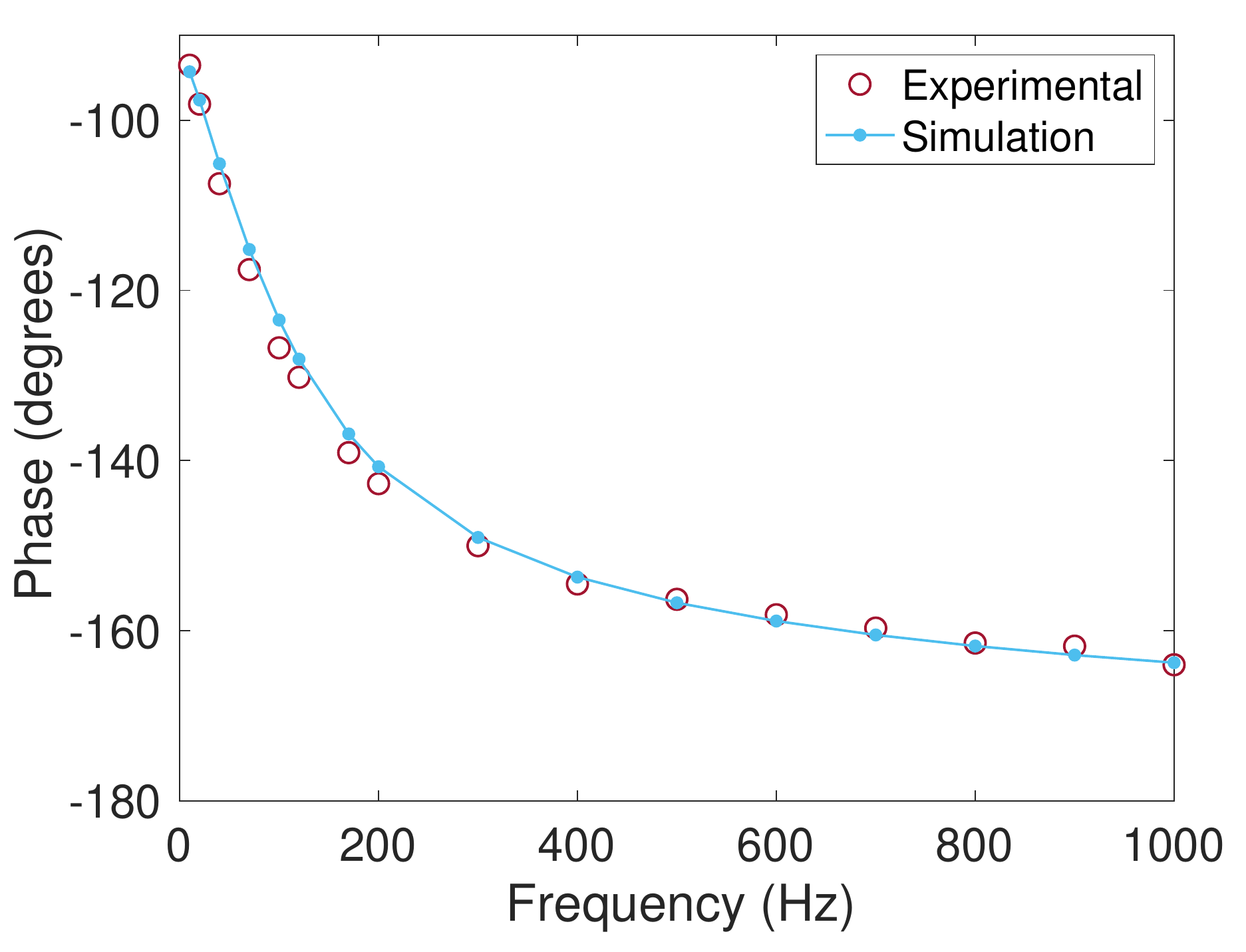}
\caption{}
\label{fig:Lucy_Freq_Solid_Alu_phase}
\end{subfigure}
\begin{subfigure}[b]{0.49\linewidth}
\centering
\includegraphics[width=\linewidth]{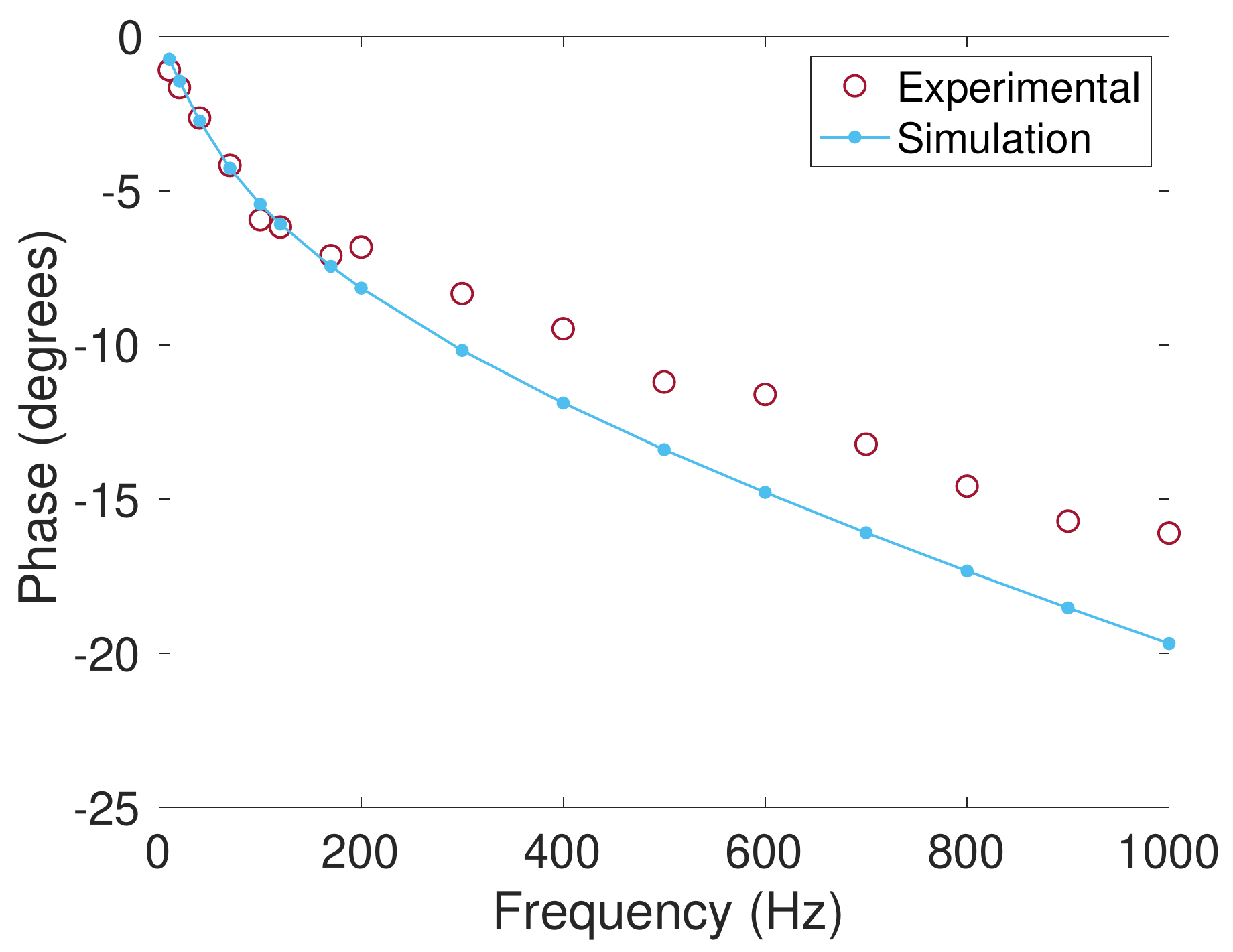}
\caption{}
\label{fig:Lucy_Freq_Solid_Steel_phase}
\end{subfigure}
\vspace{0.2cm}
\caption{Experimental and simulation results for the secondary magnetic field for solid aluminium ((a),(c) and (e)) and steel cylinders ((b),(d) and (f)) when varying the frequency between 10~Hz to 1~kHz. (a), (b): In-phase $I$ and out-of-phase $Q$ components. (c), (d): ratio of the amplitude of the secondary magnetic field to the primary magnetic field at the magnetometer position. (e), (f): Phase (in degrees) of the secondary magnetic field with respect to the primary magnetic field. 
}
\label{fig:Lucy_Freq}
\end{figure}

Figure~\ref{fig:Lucy_Freq} shows the detected secondary magnetic field for solid aluminium (Figs.~\ref{fig:Lucy_Freq_Solid_Alu_I/Q}, \ref{fig:Lucy_Freq_Solid_Alu_B_ratio}, \ref{fig:Lucy_Freq_Solid_Alu_phase})
and steel (Figs.~\ref{fig:Lucy_Freq_Solid_Steel_I/Q}, \ref{fig:Lucy_Freq_Solid_Steel_B_ratio}, \ref{fig:Lucy_Freq_Solid_Steel_phase})
cylinders as a function of the excitation frequency $f$. 
The figure shows the in-phase $\Delta I$ and out-of phase $\Delta Q$ components of the secondary field normalised to the amplitude of the primary field 
(Figs.~\ref{fig:Lucy_Freq_Solid_Alu_I/Q} and \ref{fig:Lucy_Freq_Solid_Steel_I/Q}), 
the normalised amplitude of the secondary field 
(Figs.~\ref{fig:Lucy_Freq_Solid_Alu_B_ratio} and \ref{fig:Lucy_Freq_Solid_Steel_B_ratio}) and the phase $\phi = \tan^{-1}(\Delta Q/\Delta I)$ of the secondary field (relative to the primary field)
(Figs.~\ref{fig:Lucy_Freq_Solid_Alu_phase} and \ref{fig:Lucy_Freq_Solid_Steel_phase}).

For the aluminium sample, the out-of-phase component $\Delta Q$ is linear up to around 50~Hz and dominates (i.e. is larger than $\Delta I$) up to $150$~Hz (see Fig.~\ref{fig:Lucy_Freq_Solid_Alu_I/Q}). The overall magnetic field ratio saturates at $\sim$ 350~Hz (see Fig.~\ref{fig:Lucy_Freq_Solid_Alu_B_ratio}) due to the skin effect. The skin effect starts to matter when the skin depth $\delta=1/\sqrt{\pi \mu \sigma f }$ becomes comparable to or smaller than the thickness $t$ of the object, 
corresponding to frequencies $f \geq 1/\left(t^{2} \pi \mu \sigma \right) = 26$~Hz, using $t$~=~2~cm, $\mu=\mu_{0}$ and $\sigma$~=~24.6~MS/m for our aluminium sample. 
For a non-magnetic sample such as our aluminium sample, the phase $|\phi| \sim 90^{\circ}$ at low frequencies, when the signal is mainly in the out-of-phase component, and approaches $|\phi| \sim 180^\circ$ at higher frequencies, when the in-phase component dominates (see Fig.~\ref{fig:Lucy_Freq_Solid_Alu_phase}). This is due to the secondary field being in the opposite direction to the primary field at high frequencies. 
The conductivity of our aluminium sample can be extracted from the gradient $\left(|B_{\mathrm{ec}}|/|B_1|\right)/f$ at low frequencies. This is done by comparing the experimentally found gradient to an analytical formula valid for a non-magnetic conductive cylinder. Detailed calculations which are based on Ref.~\cite{griffiths1999magnetic} are presented in Appendix~\ref{alu_cond}.
Using the gradient of the magnetic field ratio, up to 20~Hz, we determine the conductivity to be $25.5~(\pm 1.8)$~MS/m, which is in agreement with the expected value of 24.6~MS/m for 6061~T6 aluminium. 
Figures~\ref{fig:Lucy_Freq_Solid_Alu_I/Q}, \ref{fig:Lucy_Freq_Solid_Alu_B_ratio} and \ref{fig:Lucy_Freq_Solid_Alu_phase} also show the results of numerical simulations carried out in COMSOL which agree very well with the experimental results.

\newpage

For our steel sample, the biggest signal is seen at low frequencies, as shown in Fig.~\ref{fig:Lucy_Freq_Solid_Steel_I/Q}. This is because steel has a magnetic permeability \cite{davis1994stainless}. The secondary field is produced in the same direction as the primary field due to steel being ferromagnetic. The in-phase component dominates at all frequencies, so as expected the phase of the signal is small $|\phi| < 20^\circ$ (Fig.~\ref{fig:Lucy_Freq_Solid_Steel_phase}) and the magnetic field ratio (Fig.~\ref{fig:Lucy_Freq_Solid_Steel_B_ratio}) is very similar to that of the in-phase component. The overall magnetic field that is detected decreases with frequency slightly but still remains a large signal. It can also be seen that at higher frequencies the out of-phase component increases as the in-phase component decreases.

The exact value of the magnetic permeability of the 440c steel samples we used was not known to us in advance. 
In \cite{honke2018metallic} their 440c steel samples were found to have  $\mu_{r}=16-17$. However, no other literature could be found where the magnetic permeability of 440c steel is calculated, so it is unknown how much this changes between samples. For low permeabilities, a small change in the permeability can cause a large change in the signal detected \cite{bidinosti2007sphere, honke2018metallic}. As $\mu_r \gg 1$ the change in the signal is a lot smaller.
Hence for a simulation comparison to be done the conductivity and permeability needed to be determined experimentally. 
In order to determine these values, we fitted our experimental results to analytical formulae from Ref.~\cite{honke2018metallic}. As those formulae are valid for a sphere in a uniform field, and experimentally our object is a cylinder and is not in a uniform RF field, we included a scale factor in the fit function (see Appendix~\ref{steel_cond}). 
For our 440c steel sample a permeability of $\mu_r = 50 ~(\pm 15)$ and a conductivity of $\sigma = 1.67~(\pm 0.2)$~MS/m were obtained from the fit (see Fig.~\ref{fig:comparison}) and then used in the simulations. The simulation results for the magnetic field ratio agree within $\sim 5\%$ with the experimental data for these parameters, with both following the same trends. Hence these values are used for the simulations throughout. 

\begin{figure}[H]
\centering
\begin{subfigure}[b]{0.49\linewidth}
\centering
\includegraphics[width=\linewidth]{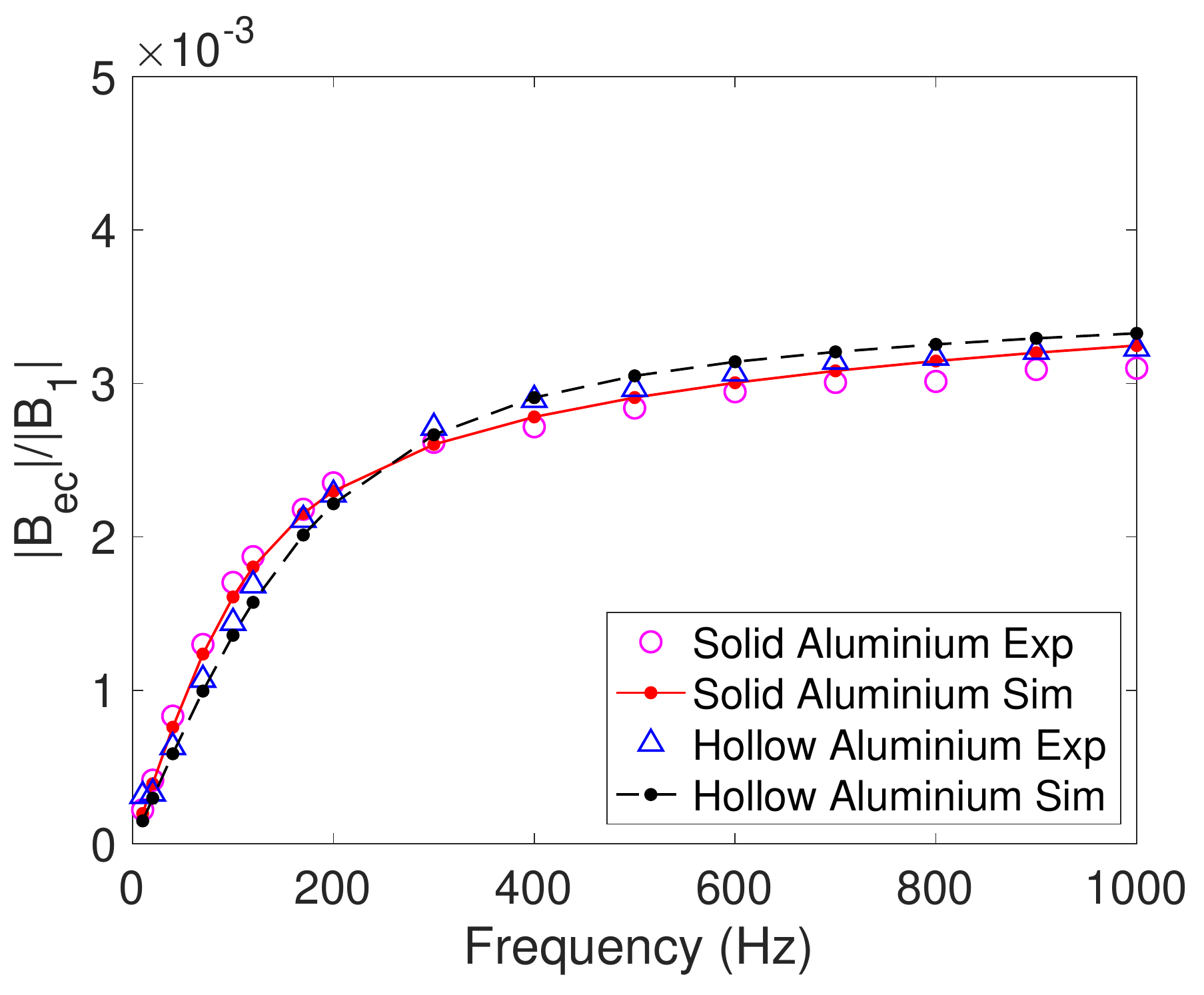}
\caption{}
\label{fig:Lucy_freq_Alu}
\end{subfigure}
\centering
\begin{subfigure}[b]{0.49\linewidth}
\centering
\includegraphics[width=\linewidth]{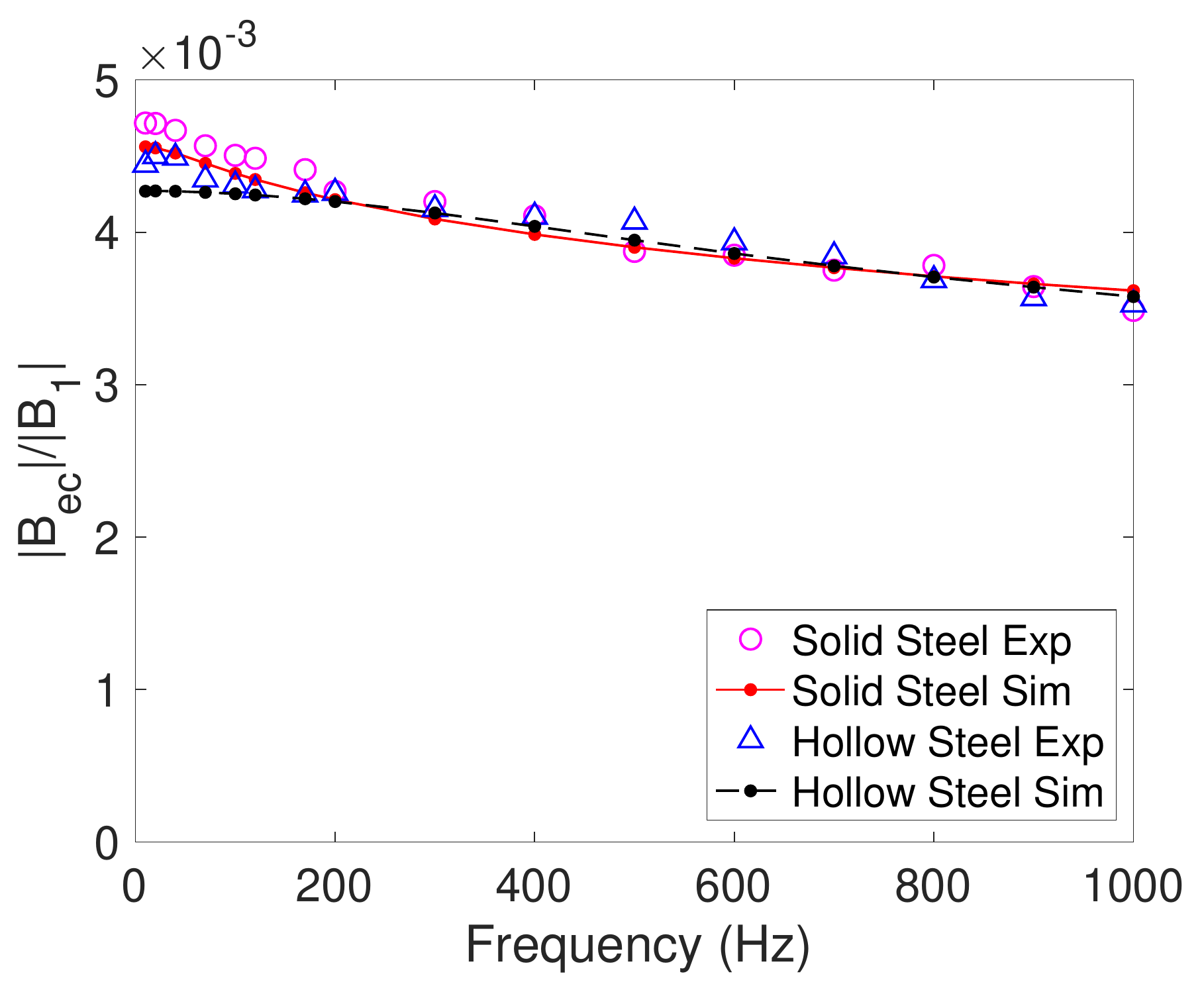}
\caption{}
\label{fig:Lucy_Freq_Steel}
\end{subfigure}
\caption{Experimental results of the secondary magnetic field from a solid and hollow cylinder for frequencies between 10 Hz and 1 kHz for (a) 6061 T6 aluminium and (b) 440c steel.}
\label{fig:_Lucy_Freq}
\end{figure}

The obtained results for the solid aluminium and steel cylinders are shown side-by-side in Fig.~\ref{fig:Lucy_Freq}. We observe that the samples can easily be differentiated by varying the excitation frequency. In particular at low-frequencies the phase of the secondary magnetic field is close to $0^\circ$ for steel (which is magnetic), while the phase is close to $90^\circ$ for aluminium (which is non-magnetic).
We also performed measurements with hollow aluminium and steel cylinders.
Figure~\ref{fig:_Lucy_Freq} shows a comparison of the magnetic field ratio $|B_{\text{ec}}|/|B_{1}|$ as a function of frequency for the solid and hollow cylinders (see Fig.~\ref{fig:Lucy_samples}). 
We find that the secondary field from the hollow cylinders is close to that of the solid cylinders. This is due to the objects having similar dimensions.

\subsection{Varying Distance}
\begin{figure}[H]
\centering
\begin{subfigure}[b]{0.49\linewidth}
\centering
\includegraphics[width=\linewidth]{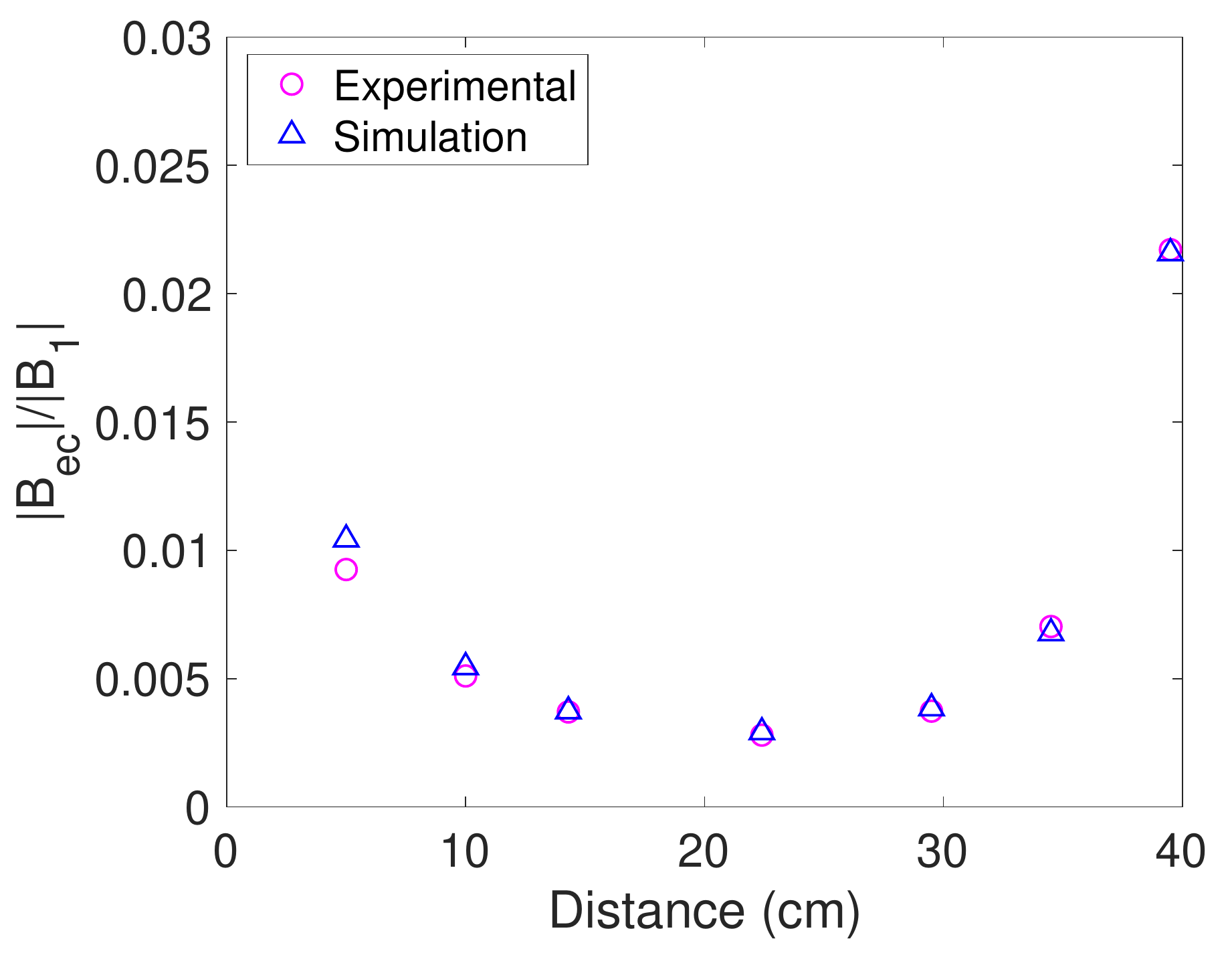}
\caption{}
\label{fig:Lucy_Dist_Alu}
\end{subfigure}
\centering
\begin{subfigure}[b]{0.49\linewidth}
\centering
\includegraphics[width=\linewidth]{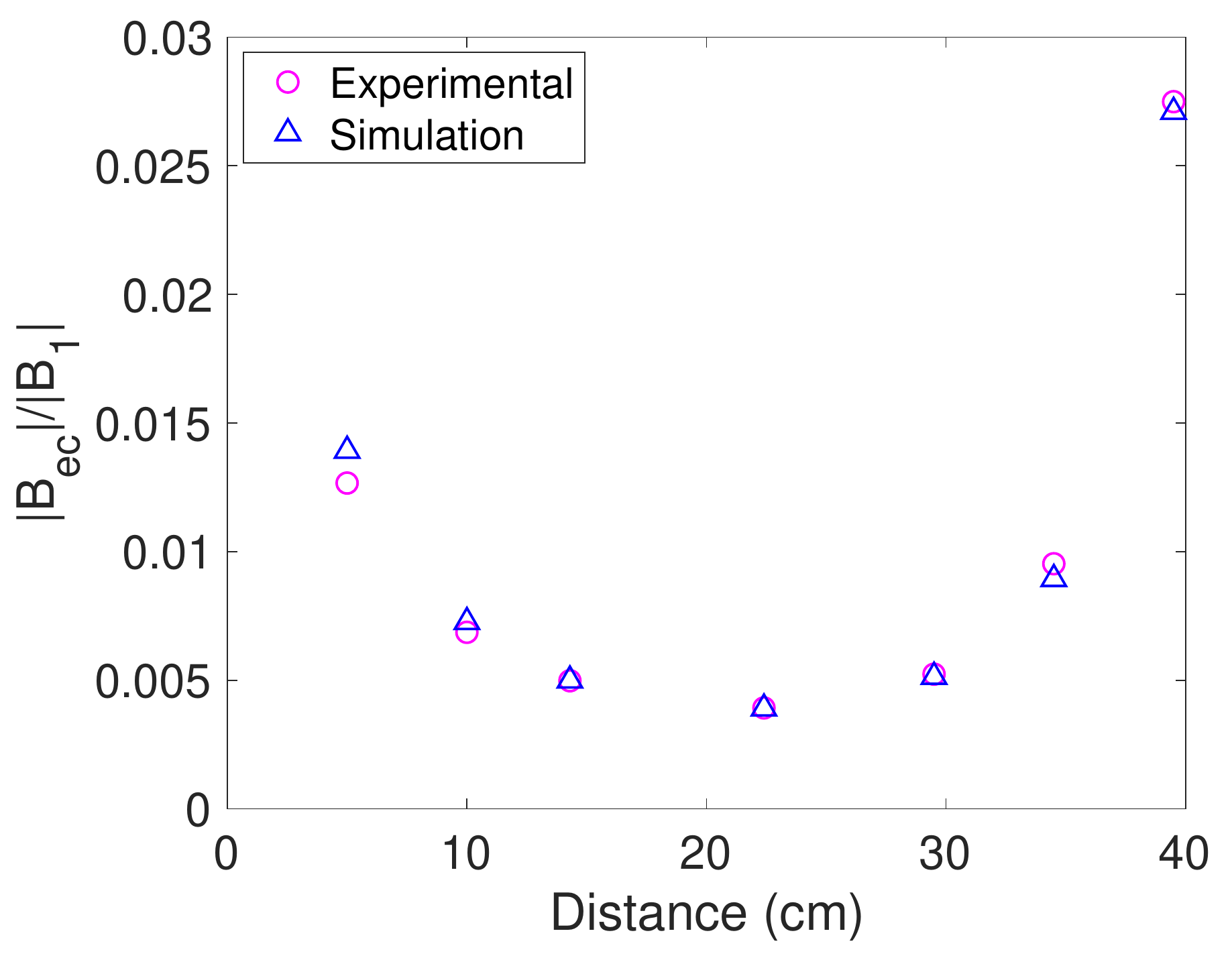}
\caption{}
\label{fig:Lucy_Dist_Steel}
\end{subfigure}
\caption{Magnetic field ratio detected as the distance from the excitation coil to the object is varied at 500 Hz for (a) the solid aluminium cylinder and (b) the solid steel cylinder.} 
\label{fig:_Lucy_Dist}
\end{figure}

We now present our on-axis measurements where the objects are placed directly in between the excitation coil and the fluxgate magnetometer.
The excitation coil is placed at 0~cm. The conductive objects are then placed at intervals between 5~cm and 39.5~cm away from the excitation coil. 
The measurements are taken at $500$~Hz. This frequency is chosen from Fig.~\ref{fig:_Lucy_Freq} as there is a large signal for both aluminium and steel.
For both aluminium and steel objects, the magnetic field ratio is smallest when the object is halfway between the excitation coil and fluxgate magnetometer and largest when the object is placed near either the excitation coil or magnetometer (see Fig.~\ref{fig:_Lucy_Dist}). By comparing Fig.~\ref{fig:Lucy_Dist_Alu} and Fig.~\ref{fig:Lucy_Dist_Steel} it can be seen that the signal is larger for steel than aluminium, as is the case when the frequency is varied. At all positions the sample can be detected. 
Figure.~\ref{fig:_Lucy_Dist} also shows the results of numerical simulations, and we again find a good agreement between experiments and simulation.

\subsection{Off-Axis Measurements}

\begin{figure}[H]
\centering
\begin{subfigure}[b]{0.49\linewidth}
\centering
\includegraphics[width=\linewidth]{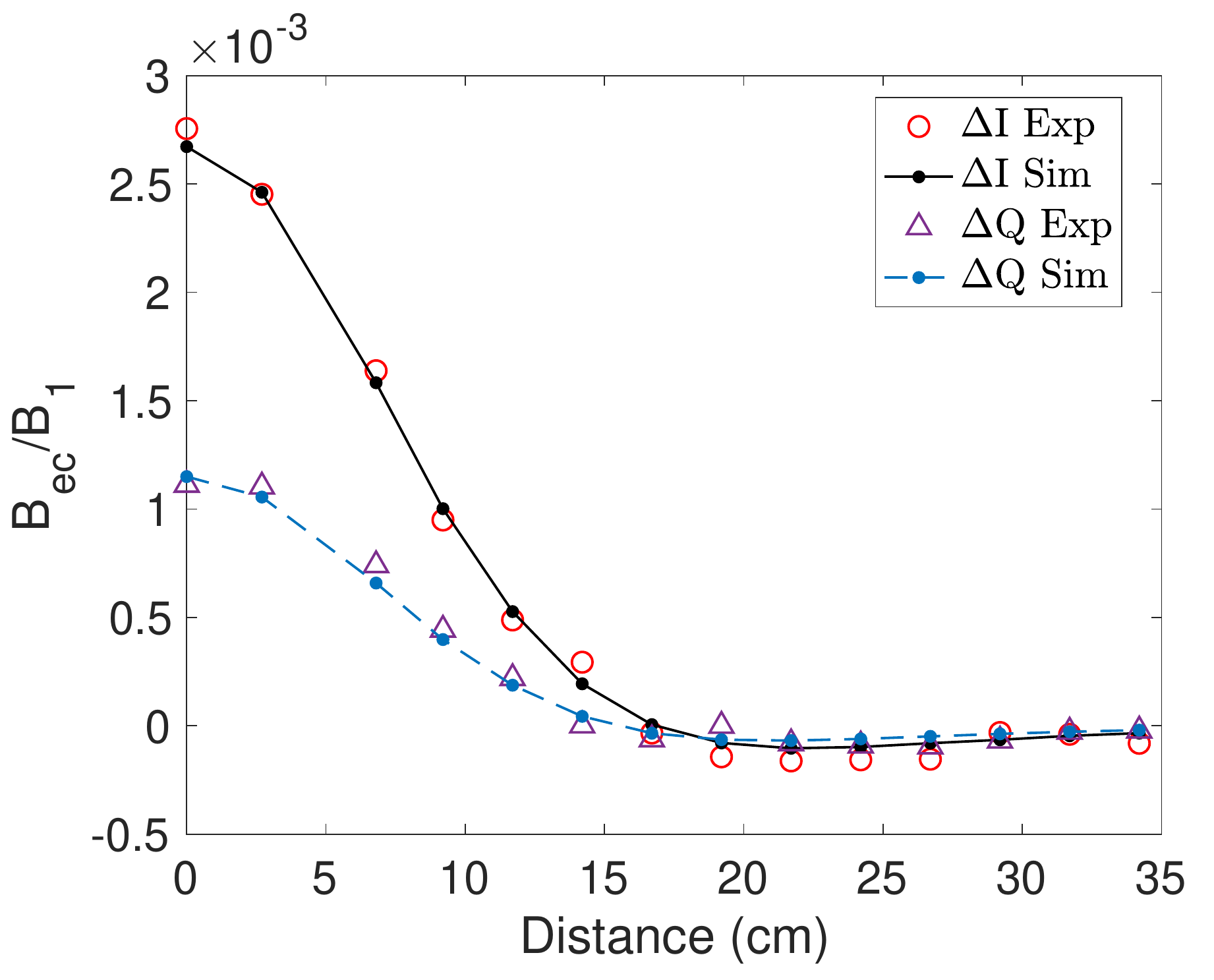}
\caption{}
\label{fig:Lucy_off_Solid_Alu_I/Q}
\end{subfigure}
\centering
\begin{subfigure}[b]{0.49\linewidth}
\centering
\includegraphics[width=\linewidth]{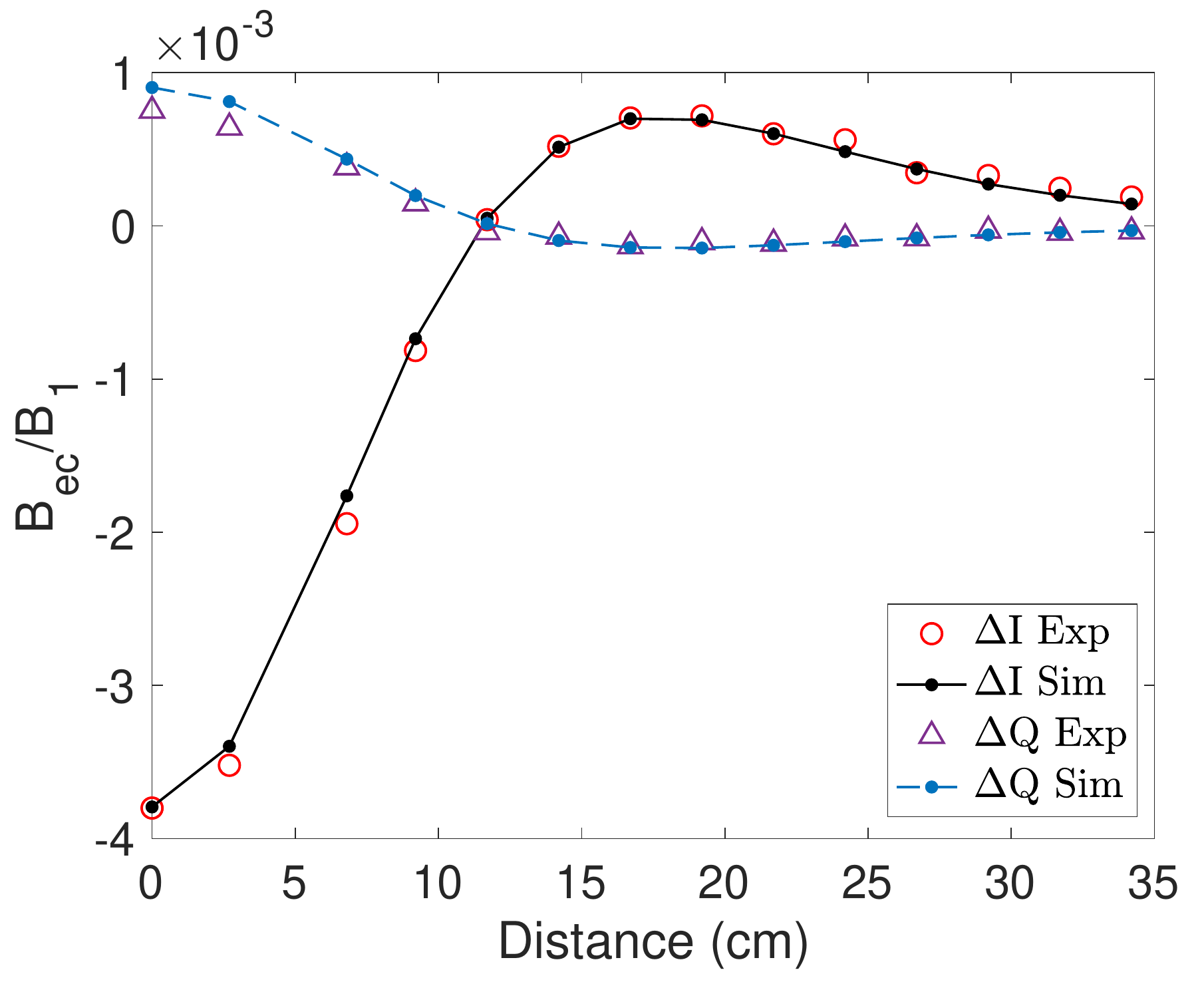}
\caption{}
\label{fig:Lucy_off_Solid_Steel_I/Q}
\end{subfigure}
\caption{In-phase and out-of-phase components of the signal detected as the conductive objects are varied off-axis at 500 Hz for the (a) solid aluminium cylinder and (b) solid steel cylinder.}
\label{fig:Lucy_off_I/Q}
\end{figure}

\begin{figure}[H]
\centering
\begin{subfigure}[b]{0.49\linewidth}
\centering
\includegraphics[width=\linewidth]{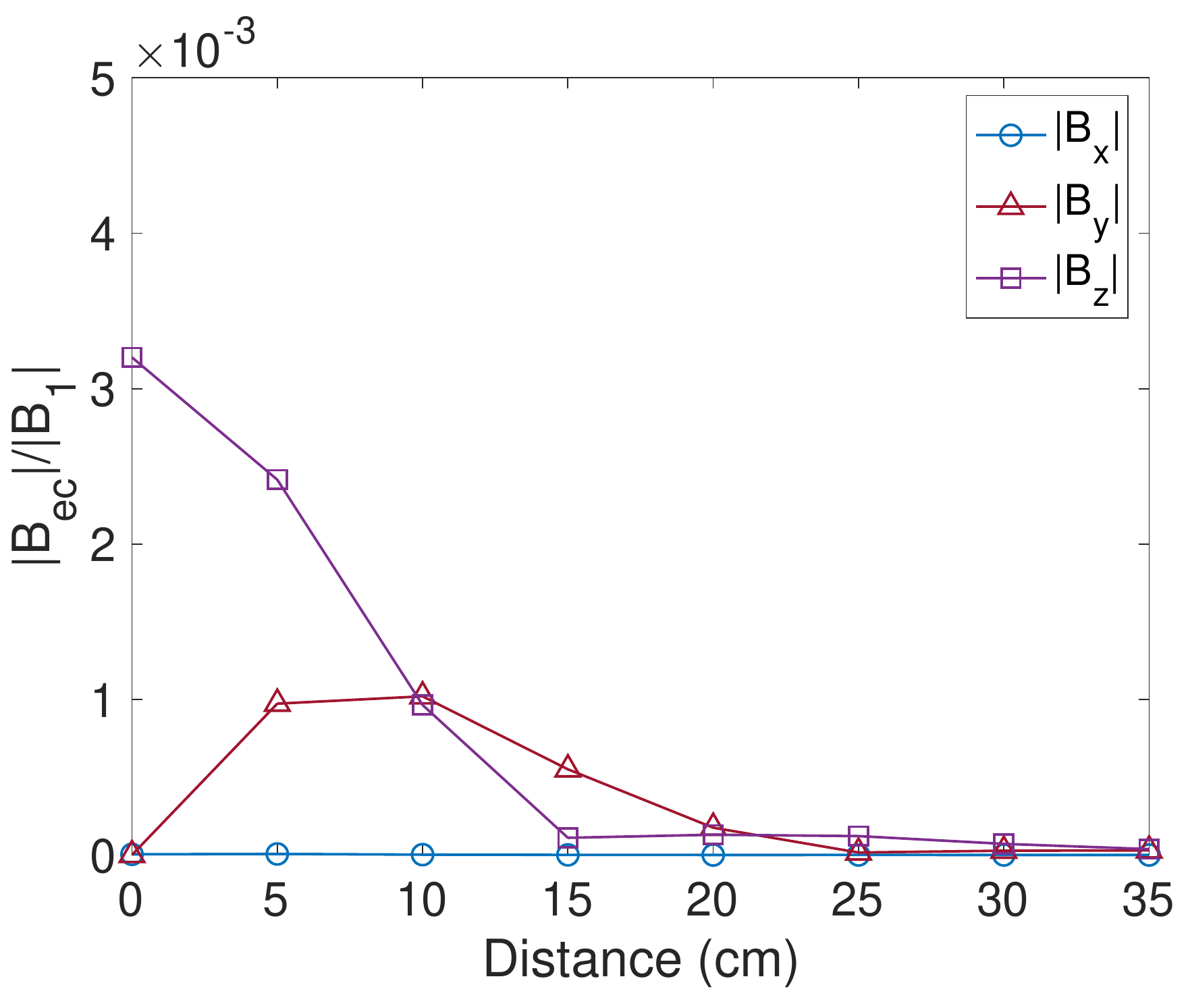}
\caption{}
\label{fig:Bec_B1}
\end{subfigure}
\centering
\begin{subfigure}[b]{0.49\linewidth}
\centering
\includegraphics[width=\linewidth]{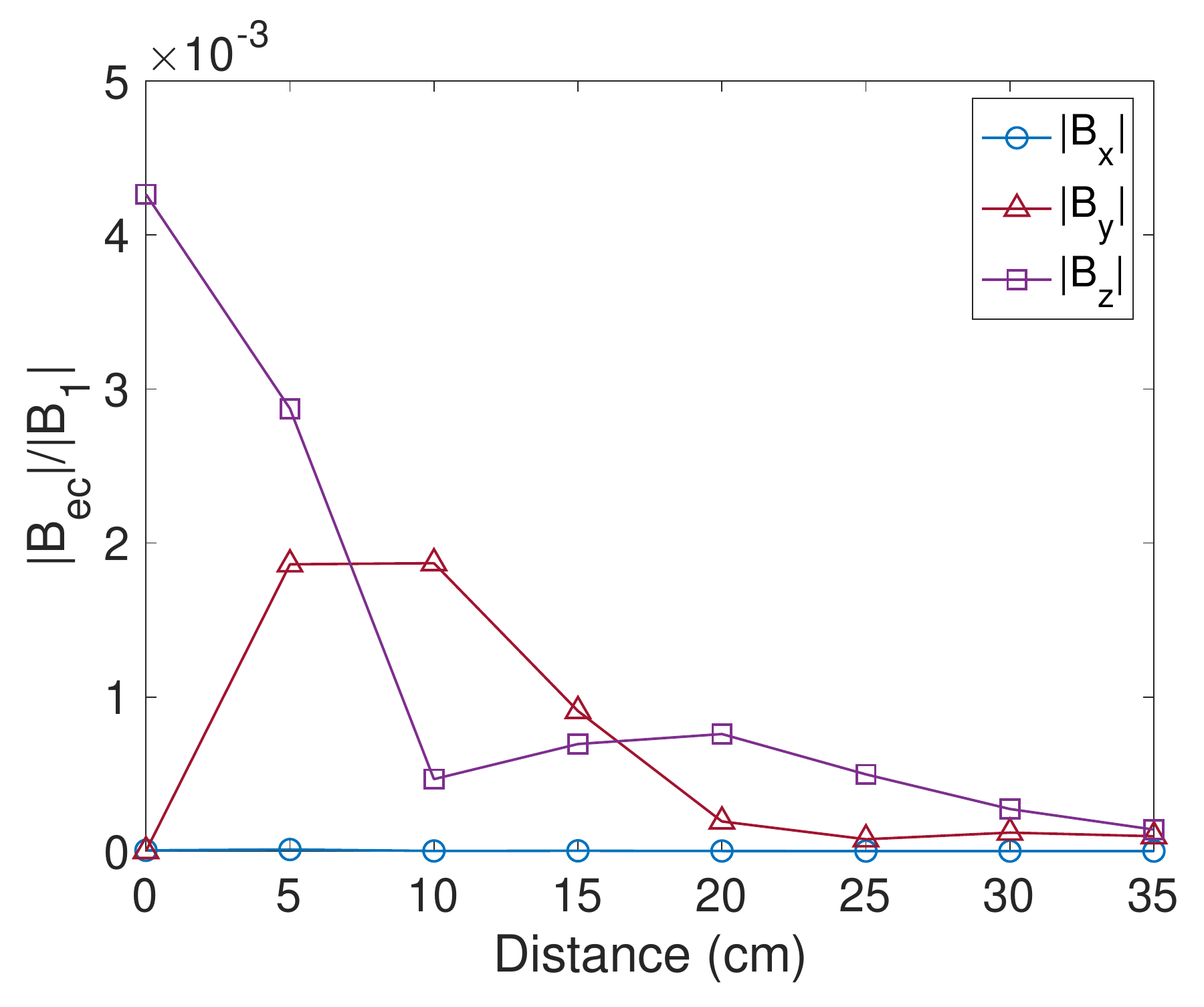}
\caption{}
\label{fig:Bec_B1_s}
\end{subfigure}
\caption{Simulation of the induced fields in the $x-$, $y-$ and $z-$direction as the (a) solid aluminium cylinder and (b) solid steel cylinder are moved off axis.}
\label{fig:Bec}
\end{figure}

We now present our results of detecting objects placed off-axis.
The object is placed at $y=0$ cm (on-axis) to $34.5$~cm off-axis. As both the aluminium and steel cylinders are moved off-axis, the in-phase and out-of-phase components get smaller and change sign, as shown in Fig.~\ref{fig:Lucy_off_I/Q}. For aluminium, the signals change sign when the object is around $16$~cm off-axis. Similarly, for steel the signal changes sign when the object is around  $12$~cm off-axis. These results are validated by  COMSOL simulations. 
The reason for the change in sign is due to the orientation of the induced dipole (see Fig.~\ref{fig:eddy_current}). In the experiment only the $z$-component of the magnetic field is recorded, however the secondary magnetic field will in general have both $y$- and $z$-components when the object is placed off-axis. To study how the vector components of the detected secondary magnetic field change  as the object is moved off axis we carried out COMSOL simulations where the object is swept along the $y$-axis (see Fig.~\ref{fig:Bec}). For the analysis of solid aluminium and steel cylinders, $B_{\text{ec,x}}$ and $B_{\text{ec,y}}$ were obtained in addition to $B_{\text{ec,z}}$. The signal is measured at an excitation frequency of 500~Hz, in order to match the experimental conditions. 
We find that the $z$-component of the secondary magnetic field is maximal when the object is on-axis (at $y=0$) and that the $y$-component reaches a maximum at around 5~cm to 10~cm for both aluminum and steel samples. The $x$-component of the field is zero as the induced dipole is in the $y$-$z$-plane.

\section{Conclusions}
In conclusion, we have detected and characterised non-magnetic (aluminium) and magnetic (steel) samples by inducing eddy currents in them and detecting the secondary magnetic field with a fluxgate magnetometer. We have shown that the samples can be differentiated by varying the frequency of the primary magnetic field. Their electrical conductivities and magnetic permeabilities were determined by fitting the experimentally measured secondary field to analytical formulae. Overall, our experimental results are in good agreement with numerical simulations carried out in COMSOL. By varying the position of the sample with respect to the excitation coil and magnetometer, we demonstrate the possibility of locating metallic objects based on the $x$-, $y$-, $z$-components of the secondary magnetic field.
Localisation of a magnetic dipole can be done using a small array of vector fluxgate magnetometers \cite{Munschy2007}. Using a primary field and detecting the induced secondary magnetic field has the advantages that both magnetic and non-magnetic objects can be detected and that the method is not sensitive to the background Earth field and the method finds applications in detection of unexploded ordnance.
The localisation and characterisation of samples could also be further explored with the help of machine learning \cite{Deans2018prl}.
It is worth noting that although we use a fluxgate magnetometer to detect eddy currents, other types of sensors can also be used, such as optically pumped magnetometers \cite{Budker2007, bevington_gartman_chalupczak_2019,bevington_gartman_chalupczak_2021, deans_cohen_yao_maddox_vigilante_renzoni_2021} or magnetoresistive sensors \cite{jander2005magnetoresistive, rifai2016giant}. Using a highly sensitive optically pumped magnetometer instead of a fluxgate magnetometer could potentially extend the detection range \cite{rushton2022unshielded}.

\authorcontributions{
Conceptualization, L.E., A.M., L.M.R., T.P. and K.J.; 
methodology,  L.E., A.M., L.M.R., T.P. and K.J.; 
software, T.P., A.M. and L.E.; 
validation, L.E., A.M., L.M.R. and K.J.; 
formal analysis, L.E., A.M., L.M.R. and K.J.; 
investigation, L.E., A.M., L.M.R., T.P. and K.J.; 
resources, L.E., A.M. and K.J.; 
data curation, L.E. and A.M.; 
writing---original draft preparation, L.E. and A.M.; 
writing---review and editing, L.E., A.M., L.M.R., T.P. and K.J..; 
visualization, L.E. and A.M.; 
supervision, K.J.; 
project administration, K.J; 
funding acquisition, T.P. and K.J. 
All authors have read and agreed to the published version of the manuscript.
}

\funding{This research was supported by the UK Quantum Technology Hub in Sensing and Timing, funded by the Engineering and Physical Sciences Research Council (EPSRC) (Grant No. EP/T001046/1), the QuantERA grant C’MON-QSENS! by EPSRC (Grant No. EP/T027126/1), the Novo Nordisk Foundation (Grant No. NNF20OC0064182), and Dstl via the Defence and Security Accelerator (www.gov.uk/dasa).}

\dataavailability{
Further data can be available from the authors upon request.}

\conflictsofinterest{The authors declare no conflict of interest.} 


\appendixtitles{yes} 
\appendixstart
\appendix
\section[\appendixname~\thesection]{Allan Deviation and Sensitivity} \label{appendix:Allan}

\begin{figure}[H]
\centering
\begin{subfigure}[b]{0.49\linewidth}
\centering
\includegraphics[width=\linewidth]{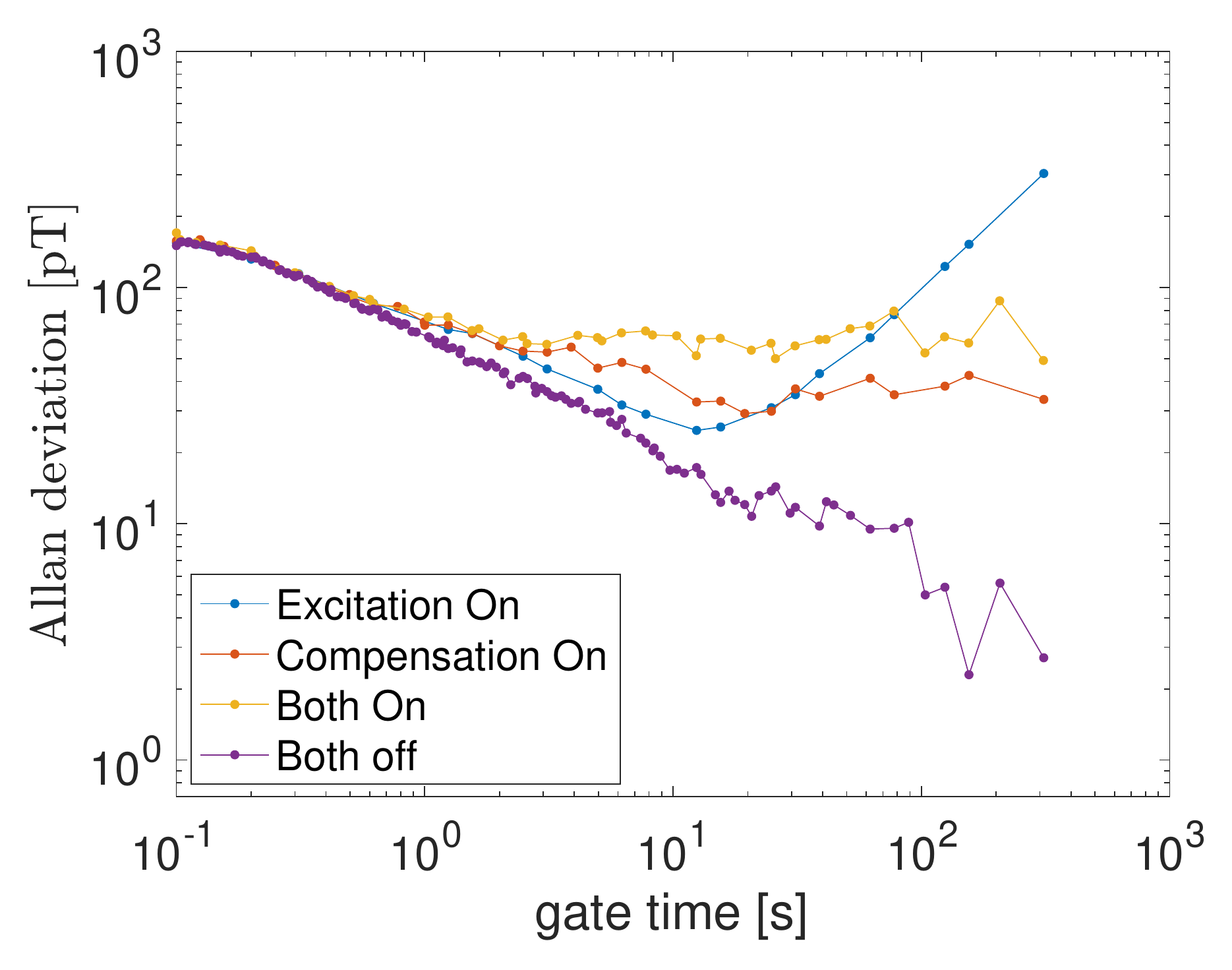}
\caption{}
\label{fig:Lucy_10_AD}
\end{subfigure}
\centering
\begin{subfigure}[b]{0.49\linewidth}
\centering
\includegraphics[width=\linewidth]{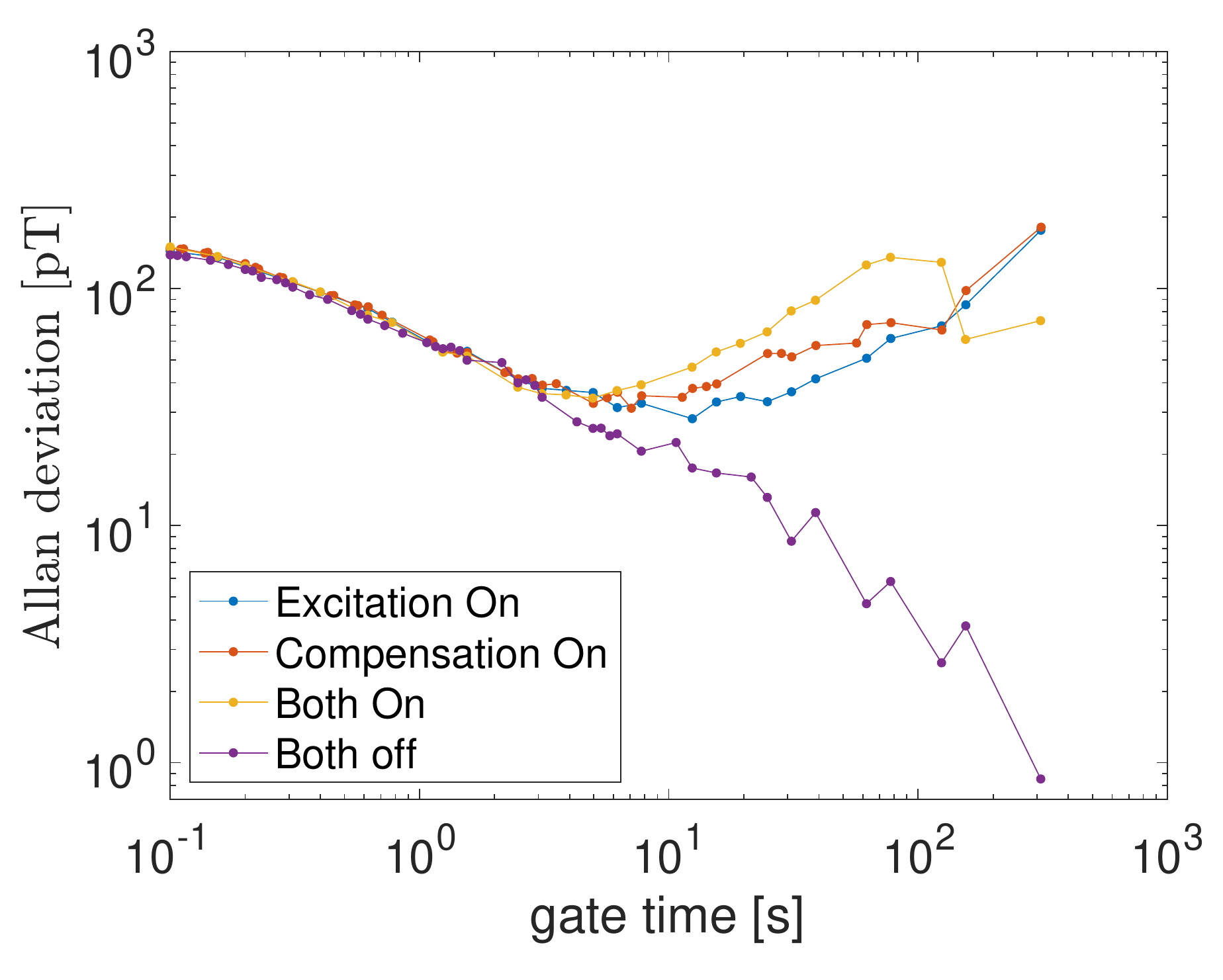}
\caption{}
\label{fig:Lucy_120_AD}
\end{subfigure}
\centering
\begin{subfigure}[b]{0.49\linewidth}
\vspace{0.2cm}
\centering
\includegraphics[width=\linewidth]{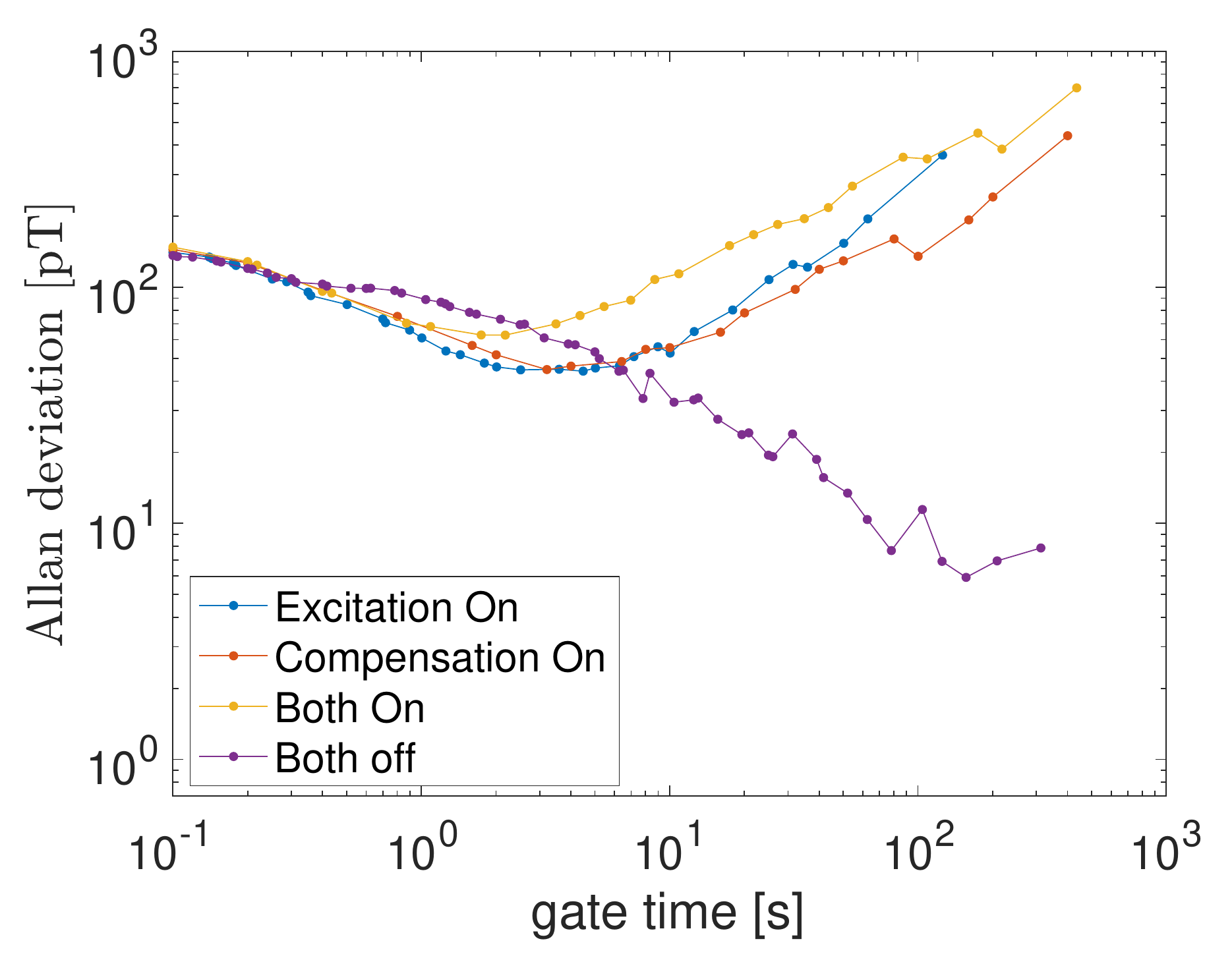}
\caption{}
\label{fig:Lucy_500_AD}
\end{subfigure}
\centering
\begin{subfigure}[b]{0.49\linewidth}
\centering
\includegraphics[width=\linewidth]{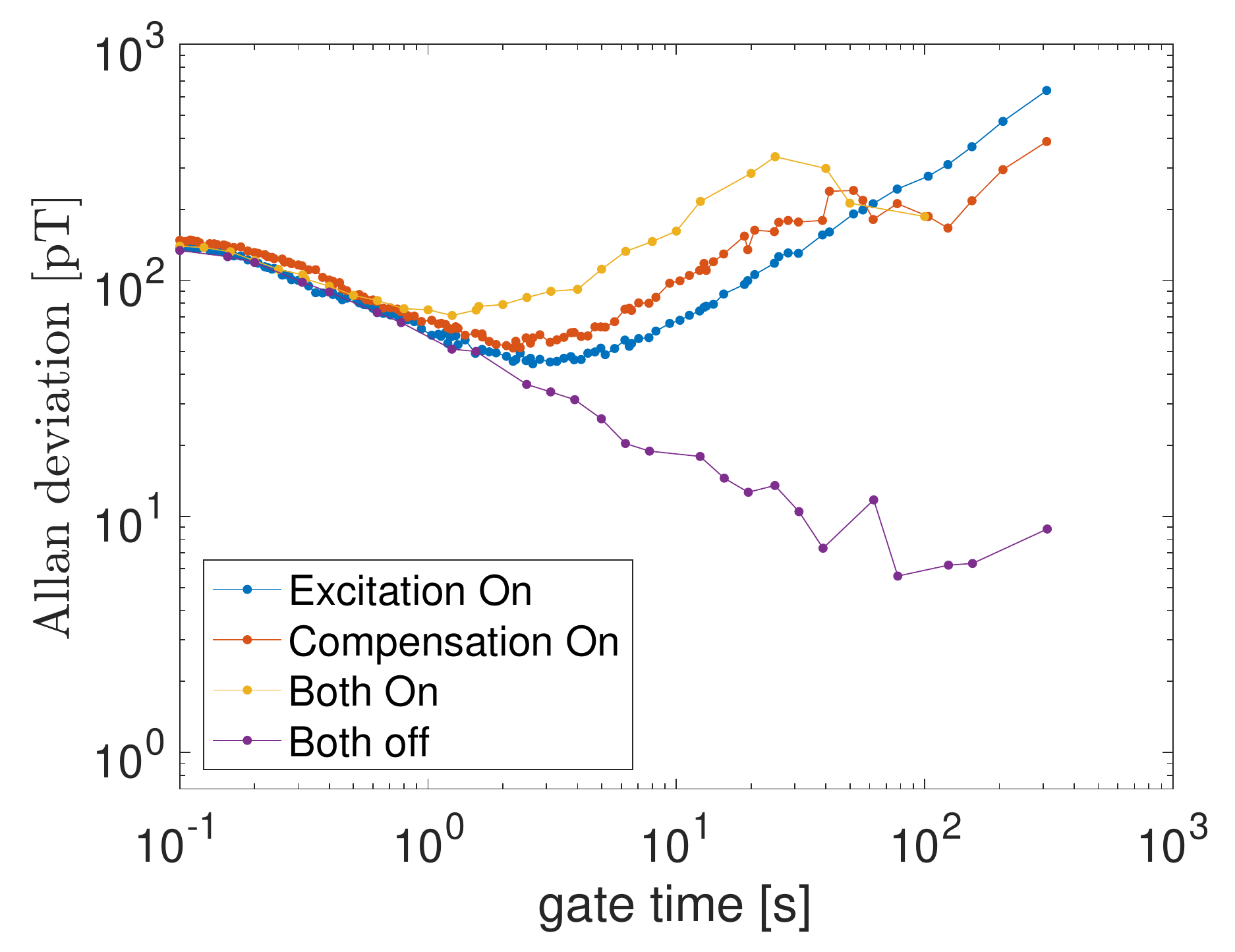}
\caption{}
\label{fig:Lucy_1k_AD}
\end{subfigure}
\caption{Allan deviation of 10 minute (unshielded) noise measurements with (i) excitation coil on, (ii) compensation coil on, (iii)  both coils on and (iv) no coils on at (a) 10 Hz, (b) 120 Hz, (c) 500 Hz and (d) 1000 Hz.}
\label{fig:Lucy_AD}
\end{figure}

In order to study the stability and sensitivity of the active detection setup, data is collected for $10$~minutes without any object at a number of different frequencies ($10$~Hz, $120$~Hz, $500$~Hz and $1$~kHz). The data was taken with a Bartington MAG690 fluxgate magnetometer connected to a FPGA which performed lock-in detection. The Allan deviation, which is a measure of frequency stability, of these noise measurements can then be calculated. The measurements performed to calculate the Allan deviation were all taken in unshielded conditions. Figure~\ref{fig:Lucy_AD} shows the Allan deviation of the in-phase component of these noise signals. The Allan deviation of the out-of-phase component is found to be similar to that of the in-phase component. In these figures it can be seen that having both coils on does not make the signal more unstable than just having one of them on. In the active detection setup both coils are used to cancel the background fields in the lab. When both coils are on the smallest detectable fields (the smallest Allan deviation) are $51$~pT at $10$~Hz, $35$~pT at $120$~Hz, $62$~pT at $500$~Hz and $70$~pT at $1$~kHz which occur at gate times $\tau \sim 1-10$~s. In Figure~\ref{fig:Lucy_AD} the noise of the fluxgate with both coils off can be understood - the Allan deviation decreases across all gate times. When the other coils are turned on, individually or together, the Allan deviation initially decreases and then increases again after a certain amount of time. This is most likely due to drifts in the current through the coils. In order to improve the stability of the coils, a less noisy current supply could be used. 

\begin{figure}[H]
\centering
\begin{subfigure}[b]{0.49\linewidth}
\centering
\includegraphics[width=\linewidth]{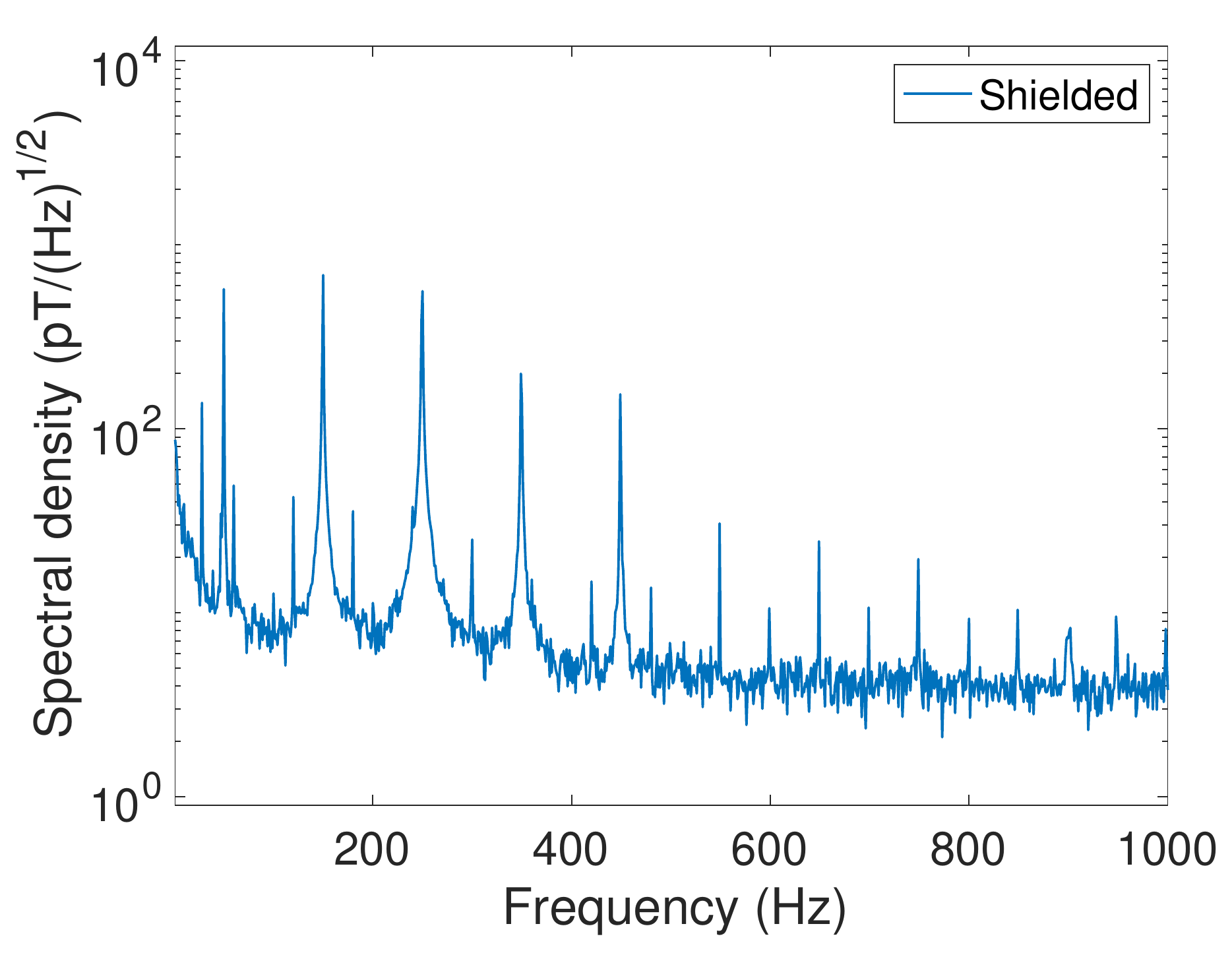}
\caption{}
\label{fig:Lucy_shielded_sensitivity}
\end{subfigure}
\centering
\begin{subfigure}[b]{0.49\linewidth}
\centering
\includegraphics[width=\linewidth]{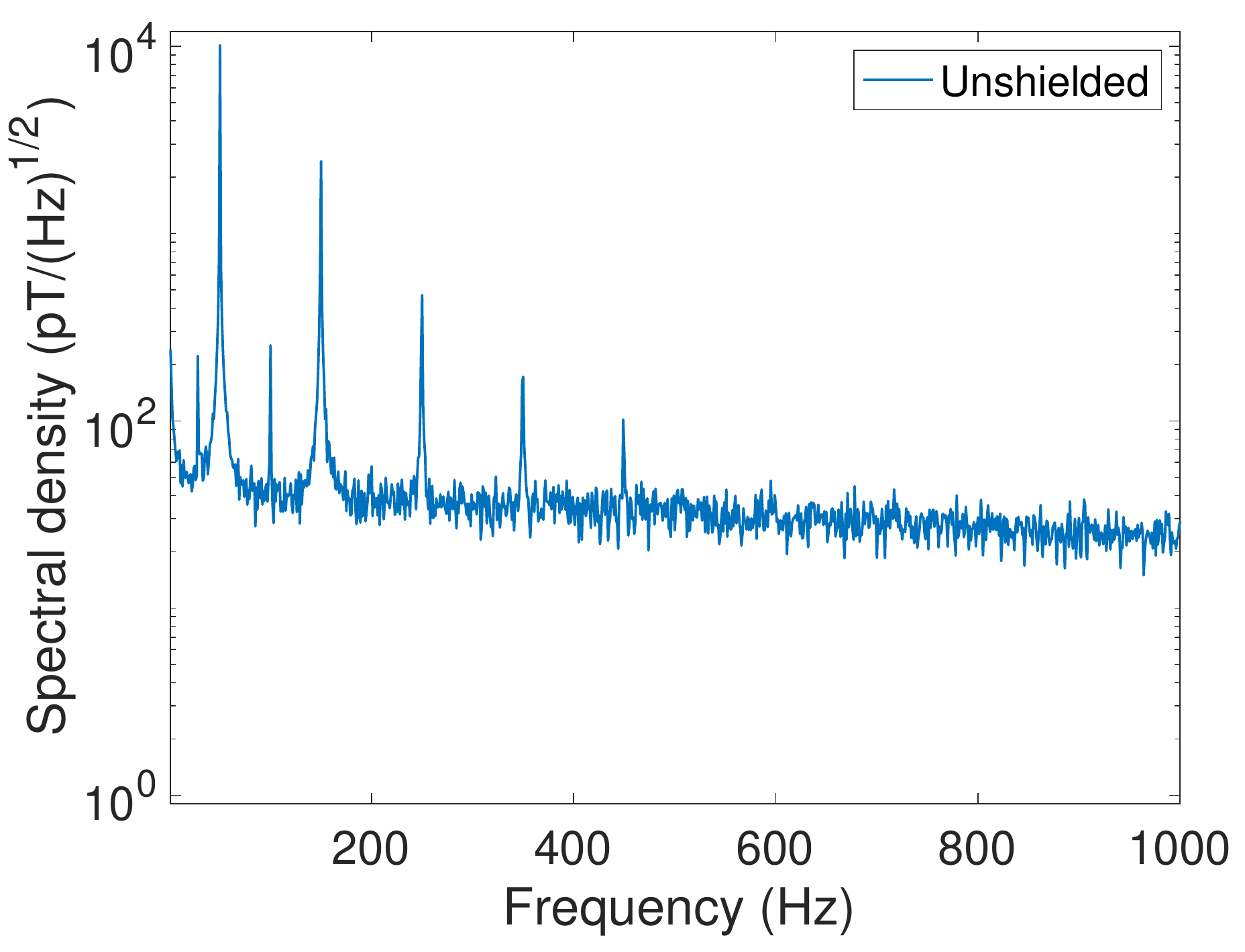}
\caption{}
\label{fig:Lucy_unshielded_sensitivity}
\end{subfigure}
\caption{Fourier transforms of 1 second time traces taken using the Bartington MAG690 magnetometer in (a) shielded and (b) unshielded conditions.}
\label{fig:Lucy_Sensitivity}
\end{figure}

In order to find the sensitivity of the fluxgate magnetometer, one second of data was collected from the magnetometer and then the power spectral density was calculated from the time trace. This was done inside a magnetic shield with all end caps on (shielded conditions) as shown in Fig.~\ref{fig:Lucy_shielded_sensitivity}, and in unshielded conditions as shown in Fig.~\ref{fig:Lucy_unshielded_sensitivity}. By taking these measurements in shielded conditions the intrinsic sensitivity of the fluxgate could be found. From Fig.~\ref{fig:Lucy_Sensitivity} it can be seen that the environmental noise at all frequencies in unshielded conditions is about an order of magnitude larger than the intrinsic noise of the sensor. The setup is hence limited by the environmental noise in the lab in unshielded conditions. In Figure~\ref{fig:Lucy_unshielded_sensitivity} the sensitivity at $10$~Hz, $120$~Hz, $500$~Hz and $1000$~Hz are found to be $\sim 25$~pT/$\sqrt{\text{Hz}}$, $\sim 30$~pT/$\sqrt{\text{Hz}}$, $\sim 40$~pT/$\sqrt{\text{Hz}}$ and $\sim 50$~pT/$\sqrt{\text{Hz}}$ respectively. 

\section[\appendixname~\thesection]{Calculating the Conductivity and Permeability}

The conductivity of aluminium and steel, as well as the permeability of steel, can be determined using theory \cite{honke2018metallic, griffiths1999magnetic}. Figure~\ref{fig:Lucy_on_set_up} shows the setup with the corresponding theory parameters for a sample of radius \textit{r} and thickness \textit{t}. The distance from the centre of the excitation coil to the centre of the sample is denoted by \textit{a} and the distance from the centre of the sample to the detection point of the magnetometer is \textit{a'}.

\subsection{Aluminium}\label{alu_cond}

Using the low frequency limit of the data shown in Fig.~\ref{fig:Lucy_Freq_Solid_Alu_B_ratio}, the conductivity of the aluminium sample used can be determined. The equations in \cite{griffiths1999magnetic} are altered to match the setup used in Section~\ref{freq}. The current induced in the cylinder in a thin area from the radial distance $\rho$ to $ \rho + d\rho$ is given by 
\begin{equation}
    dI = \frac{mt}{2 \pi \delta^2} \frac{\rho}{(a^2 + \rho^2)^{3/2}} d\rho,
\end{equation}
where $m$ is the magnetic dipole moment of the coil and $\delta$ is the skin depth \cite{griffiths1999magnetic}. This current induces a magnetic field that at the detection point measures

\begin{equation}
    dB_{\text{ec}} = \frac{\mu_0 dI}{2} \frac{\rho^2}{(\rho^2 +a'^2)^{3/2}}. 
\end{equation}
Integrating from the centre of the cylinder $\rho = 0$ to the radius of the cylinder $\rho = r$ gives the total field induced by the eddy currents at the detection point of the magnetometer. For $a \ne a'$

\begin{equation}\label{bec}
    B_{\text{ec}} = \frac{m t \mu_0}{4 \pi \delta^2} \left( \frac{a^2(2a'^2 + r^2) +a'^2 r^2}{(a^2 - a'^2)^2 \sqrt{a^2 + r^2}\sqrt{a'^2 + r^2}} - \frac{2aa'}{(a^2 - a'^2)^2}\right). 
\end{equation}
Note if $a = a'$ then equation~\ref{bec} is not defined. This is due to the integral simplifying and giving a simpler equation which is shown in \cite{griffiths1999magnetic}. Both equations tend to the same limit as $a\rightarrow a'$. By substituting in the skin depth $\delta^{2} = 1/\left(f \pi \mu \sigma \right)$ and dividing through by $f B_1 = f \mu_0m /2(a+a')^3\pi$ at the detection point of the magnetometer it can be found that

\begin{equation}
    \frac{B_{\text{ec}}}{B_1 f} = \frac{t \pi \mu \sigma}{2} \frac{(a+a')^3}{(a^2-a'^2)^2} \left( \frac{a^2 (2a'^2 +r^2) +a'^2 r^2 }{\sqrt{a^2 + r^2}\sqrt{a'^2 + r^2}} -2aa'\right),
\end{equation}
where the only unknown is $\sigma$. Here the left hand side is given by the gradient of the magnetic field ratio in the low frequency limit. By using a fit function in MATLAB and the experimental data for the aluminium cylinder in Fig.~\ref{fig:_Lucy_Freq} we find a conductivity of $25.5~(\pm 1.8)$~MS/m, which is in agreement with the data sheet for 6061 T6 aluminium \cite{davis1993aluminum}.

\subsection{Steel}\label{steel_cond}

In order to calculate the conductivity and magnetic permeability of the steel sample used, the equations in \cite{honke2018metallic, bidinosti2007sphere} needed to be altered to match the setup in Fig.~\ref{fig:Lucy_on_set_up}. Using data from Section~(\ref{freq}) we can determine the conductivity and permeability of our 440c steel cylinder. 

The primary magnetic field at the centre of the cylinder is given by 
\begin{equation}\label{1}
    B_1(z=a) = \frac{\mu_0 m}{2 \pi a^{3}},
\end{equation}
where $m$ is the magnetic moment of the excitation coil, $a$ is the distance from the centre of the excitation coil to the centre of the object and $\mu_0$ is the vacuum permeability. This induces a secondary field at the position of the sensor equal to  
\begin{equation}\label{2}
    B_{\text{ec}}(z = a + a') = \frac{\mu_0 m_{\text{ec}}}{2 \pi a'^{3}},
\end{equation}
where $a$ is the distance from the centre of the object to the detection point of the fluxgate magnetometer and $m_{\text{ec}}$ is the induced magnetic moment in the object. For a sphere of radius $r$ the magnetic moment is given by 
\begin{equation}
    m_{\text{ec}} = \frac{2 \pi r^3 B_1(z = a)}{\mu_0} \frac{2 (\mu_r - 1)j_0(kr) + (2\mu_r+1)j_2(kr)}{(\mu_r+2)j_0(kr) + (\mu_r - 1)j_2(kr)},
\end{equation}
where $j_0$ and $j_2$ are spherical Bessel functions and $\mu_r$ is the relative permeability (to be determined) \cite{honke2018metallic}. The propagation constant is given by $k = \sqrt{\mu \varepsilon \omega^2 + i \mu \sigma \omega}$ where $\varepsilon$ is the permitivity of the sample, $\mu = \mu_0 \mu_r$ and $\omega = 2 \pi \nu$.

\begin{figure}[H]
    \centering
    \includegraphics[width=0.7\linewidth]{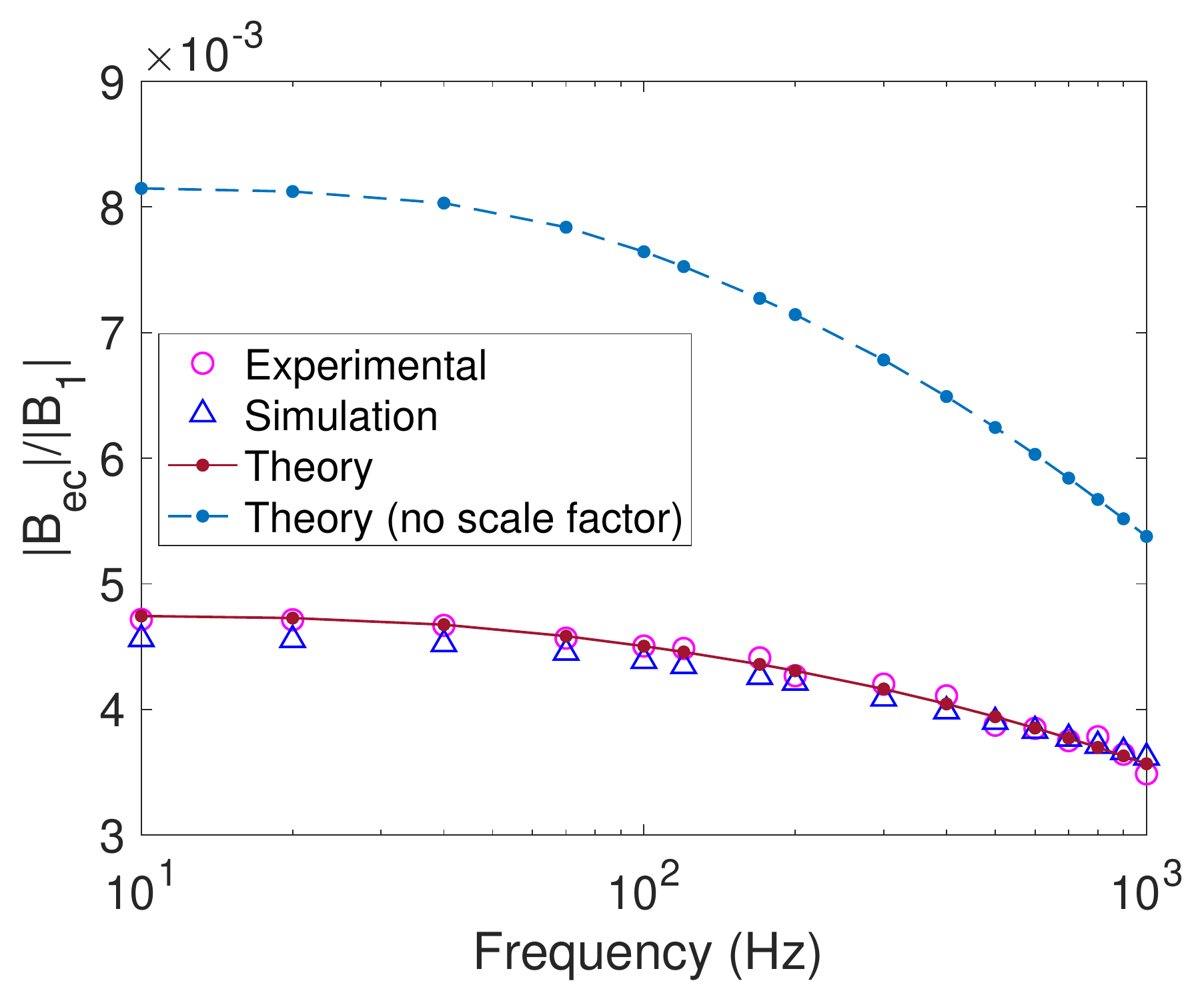}
    \caption{Fitting the conductivity and permeability of 440c steel using theory equations similar to those found in \cite{honke2018metallic}.}
    \label{fig:comparison}
\end{figure}

By combining Equations~(\ref{1}) and (\ref{2}) at the detection point of the magnetometer an equation for the magnetic field ratio can be obtained 

\begin{equation} \label{3}
    \frac{B_{\text{ec}}(z= a+a')}{B_1(z=a+a')} = \frac{r^3(a+a')^3}{(aa')^3} \frac{2(\mu_r-1)j_0(kr) + (2\mu_r + 1)j_2(kr)}{(\mu_r +1)j_0(kr) + (\mu_r -1)j_2(kr)}.
\end{equation}
This equation assumes a uniform RF magnetic field \cite{honke2018metallic} across a sphere. However, the data we collected was for a cylinder with a 2~cm radius.
As the cylinder has a finite radius the primary field across it will also not be completely uniform. 
Nonetheless, we found that our experimental results are well described by Equation~(\ref{3}) if it is multiplied by a scale factor.
Using a fit function in MATLAB, and the parameters outlined in Section~\ref{method}, we were able to determine $\mu_r$, $\sigma$ and the scale factor for 440c steel. The scale factor was calculated to be $0.56~(\pm 0.01)$, so the theory over estimated $|B_{\text{ec}}|/|B_1|$ by 79~($\pm 3$)$\%$. The values for the permeability and conductivity were calculated to be $\mu_r = 50 ~(\pm 15)$ and $\sigma = 1.67~(\pm 0.20)$~MS/m. The function was fitted to the real and imaginary components of the field ratio $|B_{\text{ec}}|/|B_1|$. This method can also be used to calculate the conductivity of aluminium as it is known that $\mu_r = 1$. In Fig.~\ref{fig:comparison} it can be seen how the theory compares with the simulation and experimental data with and without the scale factor.

\end{paracol}

\reftitle{References}

\end{document}